\documentclass[12pt]{article}

\input{epsf}
\usepackage[nosort]{cite}
\usepackage[dvips]{graphicx}
\usepackage{amsmath}
\usepackage{epsfig}
\usepackage{amsfonts}
\usepackage{amssymb}
\usepackage{multirow}
\usepackage{array}
\usepackage{epsfig}
\usepackage{amsmath}
\usepackage{amssymb}

\newcommand{\be}{\begin{equation}}
\newcommand{\ee}{\end{equation}}
\newcommand{\bee}{\begin{equation*}}
\newcommand{\eee}{\end{equation*}}
\newcommand{\bea}{\begin{eqnarray}}
\newcommand{\eea}{\end{eqnarray}}
\newcommand{\bean}{\begin{eqnarray*}}
\newcommand{\eean}{\end{eqnarray*}}

\begin{document}

\setcounter{page}{0}
\thispagestyle{empty}

\begin{flushright}
SACLAY-T11/209

ULB-TH/11-25
\end{flushright}

\vskip 8pt

\begin{center}
{\bf \LARGE {Supersonic Electroweak Baryogenesis:\\
\vskip 8pt
Achieving Baryogenesis for Fast Bubble Walls
 }}
\end{center}

\vskip 12pt

\begin{center}
{\bf   Chiara Caprini$^{a}$
  and Jos\'e M. No$^{a,b}$ }
\end{center}

\vskip 20pt

\begin{center}

\centerline{$^{a}${\it Institut de Physique Th\'eorique, CEA/Saclay, F-91191 
Gif-sur-Yvette, France}}
\centerline{$^{b}${\it Service de Physique Th\'eorique, Universit\'e Libre de Bruxelles, 
B-1050 Bruxelles, Belgium}}
\vskip .3cm
\centerline{\tt chiara.caprini@cea.fr, jose-miguel.no@cea.fr}
\end{center}

\vskip 13pt

\begin{abstract}

Standard electroweak baryogenesis in the context of a first 
order phase transition is effective in generating the baryon asymmetry of the 
universe if the broken phase bubbles expand at subsonic speed, so that CP asymmetric 
currents can diffuse in front of the wall. Here we present a new mechanism for 
electroweak baryogenesis which operates for supersonic bubble walls. It relies 
on the formation of small bubbles of the symmetric phase behind the bubble wall, in the broken 
phase, due to the heating of the plasma as the wall passes by. We apply the 
mechanism to a model in which the Higgs field is coupled to several singlets, and 
find that enough baryon asymmetry is generated for reasonable values of the parameter space.
\end{abstract}

\newpage

\tableofcontents

\section{Introduction}

The most interesting aspects of a first order electroweak (EW) phase transition are that it 
may explain the observed baryon asymmetry of the universe through EW 
baryogenesis\cite{Cohen:1990py,Cohen:1991iu,Joyce:1994fu}, and that it generates a stochastic 
background of gravitational waves (GWs) possibly detectable by next generation space-based 
interferometers \cite{Witten:1984rs,Hogan:1986,Kosowsky:1991ua,Kamionkowski:1993fg,Apreda:2001us,Grojean:2006bp}. 
The dynamics of the growth of the broken phase bubbles, and in particular the bubble wall velocity, 
play a fundamental role in both aspects. Fast walls are needed for sizable GW signals (the amplitude of the signal depending, 
among others, on the cube of the wall velocity $v_w$ \cite{Caprini:2007xq,Huber:2008hg,Caprini:2009fx}), 
while slow walls are needed for EW baryogenesis: only subsonic 
walls (or a compression wave) lead to efficient diffusion of particle asymmetries in front of the wall \cite{Joyce:1994fu}. 

An explicit computation of the bubble wall velocity has been done in the Standard Model (SM)\cite{Moore:1995si} and 
in the Minimal Supersymmetric Standard Model (MSSM)\cite{John:2000zq}, with the result that in the MSSM 
the bubbles are quite slow ($v_w \sim 0.05 - 0.1$), whereas in the SM the wall velocity is significantly larger, but still
subsonic ($v_w \sim 0.35 - 0.45$). However, for the SM the EW phase transition is 
not first order but rather a smooth cross-over \cite{Kajantie} for
values of the Higgs mass above the LEP bound $m_h > 114.4$ GeV \cite{LEP}, and MSSM EW baryogenesis is 
now strongly disfavoured by LEP and Tevatron data \cite{Germano} and may be even more challenged by the LHC in the near future. 
Yet, there are other extensions of the SM where the three Sakharov conditions for baryogenesis are naturally
satisfied during the EW phase transition, such as for example composite Higgs models \cite{EGKR}, or extensions
of the MSSM with a stronger EW phase transition \cite{Blum,Huber:2007vva}. Qualitative arguments 
show that in many of these extensions that lead to a first order 
EW phase transition, the wall velocity could be rather large, since it
increases with the strength of the phase transition \cite{Ignatius:1993qn}. This happens typically in the case of
non-supersymmetric extensions, where few degrees of freedom are added to the SM below or close to the TeV scale, 
but also in the case of supersymmetric extensions, if the first order 
phase transition is strong enough. Therefore, in these scenarios it might 
happen that the three Sakharov conditions are satisfied, yet standard EW baryogenesis cannot operate because the bubble wall 
velocity is large enough to suppress the diffusion of the particle asymmetries ahead of the bubble wall. 

In this perspective, we present here a new EW baryogenesis mechanism which operates in the case of supersonic 
bubble walls. We focus in particular on bubble expansion as a detonation. The mechanism is based 
on the fact that the expansion of the bubble as a detonation causes heating of the broken phase plasma in a small 
region behind the bubble wall \cite{EKNS}. If the temperature in this region exceeds the critical temperature $T_c$, 
bubbles of the symmetric phase can nucleate and grow inside the broken phase bubble. In these ``symmetric" bubbles 
sphalerons are reactivated, due to the restoration of the EW symmetry, and baryon number is not conserved. The expanding symmetric bubbles
move along the detonation wave towards the centre of the broken phase bubble, 
and at some point they enter the region where the heating is not 
very effective and the temperature is lower than $T_c$. In this region, the expansion of the symmetric bubbles is not any 
longer energetically favourable, and the bubbles start to shrink: the plasma then flows from the symmetric to the broken 
phase across the symmetric bubbles wall, 
and the baryon asymmetry generated inside the symmetric bubbles is transferred to the broken phase. We show that this 
process happens at a sufficiently low velocity to guarantee the diffusion of the asymmetry. 
For this mechanism to work, one further has to make sure that a large enough volume of the plasma goes 
through the symmetric phase, to explain the observed baryon asymmetry of the universe. 
We apply the mechanism to a model in which the Higgs field is coupled to several singlets, and analyse for which values 
of the parameters in the 
model (the universal coupling, the wall velocity, the Higgs mass) the filling factor is large enough and EW 
baryogenesis can occur for supersonic broken phase bubble walls. 

Having significant baryogenesis in models where the wall velocity is supersonic could accommodate a sizable GW signal with effective  baryogenesis. Clearly, scenarios which can explain the baryon asymmetry of the universe and at the same time be tested with GW detectors are 
potentially very interesting (see for example \cite{KS}). Note that this can happen also in the case of subsonic 
wall velocities: in \cite{NoGW} it has been pointed out that the relevant velocity for transport, being the velocity 
of the plasma as the wall sweeps through it, may be in general significantly lower than the wall velocity (although
a more detailed analysis is needed). 

To describe the evolution of the broken phase bubbles, we use the hydrodynamic description 
of the bubble expansion (based 
on \cite{Landau} and first applied to primordial cosmological phase transitions in \cite{Steinhardt:1981ct}), where the plasma 
around the bubble is in thermal equilibrium and the bubble wall reaches a steady state with constant 
velocity, while the center of the bubble is at rest. However, 
this picture cannot be applied to the symmetric 
bubbles, since they expand in an inhomogeneous and anisotropic background (the detonation wave of the supersonic broken phase 
bubbles). We therefore develop a qualitative description of the symmetric bubble 
expansion, which is reliable and captures the relevant aspects of the 
problem: a more detailed treatment is beyond the scope of this paper.

The paper is organised as follows: in sections \ref{sec_2} and \ref{sec_3} we review the hydrodynamics of bubble 
growth, the behaviour of the plasma as the detonation wave passes through, and the heating of the plasma in the 
rarefaction wave behind the bubble wall. In section \ref{sec_4} we demonstrate that nucleation of symmetric bubbles 
is possible, and in section \ref{sec_5} we analyse how they grow after nucleation. Then, we turn to baryogenesis: 
in section \ref{sec_6} we discuss in detail the way in which baryogenesis takes place and 
evaluate the filling factor, and in section \ref{sec_7} we analize a specific scenario 
as an example in which supersonic baryogenesis is actually possible. Finally, we conclude in section \ref{sec:Conclusions}.
Some of the relevant but involved calculations are left for the appendixes.

\section{Hydrodynamics of Bubble Growth}
\label{sec_2}

In this section we introduce the hydrodynamic 
analysis of the combined wall-plasma system\cite{Landau, Steinhardt:1981ct} (for
a recent review, see \cite{EKNS}), that will later be used in sections \ref{sec_3} and \ref{sec_5}. 
The two basic assumptions leading to the hydrodynamic approximation 
are local thermal equilibrium in the plasma and energy-momentum conservation in the 
system\footnote[2]{The system is governed by 
a set of coupled equations of motion for the plasma and the Higgs field. In particular, 
the plasma is very well described by a fluid with phase space density $f(\vec{p},\vec{x},t)$ obeying a Boltzmann equation. 
Then, when the interaction rates in 
the plasma are fast, the collision integral in the Boltzmann equation forces the phase space density towards a 
form that minimizes it, namely local thermal equilibrium, and the equations 
of motion for the plasma can be replaced by these two assumptions.}. 
The energy-momentum tensor of the Higgs field $\phi $ is given by

\be
\label{eq:TmunuHiggs}
T_{\mu\nu}^{\phi} = \partial_{\mu}\phi \partial_{\nu}\phi -g_{\mu\nu} 
\left[ \frac{1}{2} \partial_{\rho}\phi \partial^{\rho}\phi - V_0 (\phi)\right]\, ,
\ee

\noindent with $V_0 (\phi)$ being the renormalized vacuum potential. The energy-momentum tensor of a plasma locally in 
thermal equilibrium is

\be
\label{eq:Tmunu}
T_{\mu\nu}^{plasma} = w \, u_\mu u_\nu  - g_{\mu\nu} \, p \, , 
\ee

\noindent where $w$ and $p$ are the plasma enthalpy and pressure, and the quantity 
$u_\mu = (\gamma, \gamma\mathbf{v})$ is the four-velocity field of the plasma.
A constant $\phi$ background from (\ref{eq:TmunuHiggs}) contributes to the total pressure, and  
will be included in $p$ from now on. The enthalpy $w$, the entropy density
$\sigma$ and the energy density $e$ are defined by ($T$ being the temperature of the plasma)

\be
\label{pande}
w \equiv T\frac{\partial p}{\partial T}\ ,\quad
\sigma \equiv \frac{\partial p}{\partial T}\ ,\quad
e\equiv T\frac{\partial p}{\partial T} -p \, .
\ee

In the cases where the bubble expands at a constant speed, 
energy-momentum conservation $\partial^\mu T_{\mu\nu} = \partial^\mu T_{\mu\nu}^{\phi} 
+ \partial^\mu T_{\mu\nu}^{\mathrm{plasma}} = 0$ reads in
the wall frame (with the wall and fluid velocities aligned in the $z$
direction) $ \partial_z T^{zz} = \partial_z T^{z0} = 0$.
Integrating these equations across the bubble wall and
denoting the phases by subscripts $+$
(symmetric phase) and $-$ (broken phase) one obtains the matching
equations\cite{Landau,Steinhardt:1981ct}

\be
\label{eq:vvs0}
v_+ v_- = \frac{p_+  - p_-}{e_+ - e_-}\ , \quad
\frac{v_+}{ v_-} = \frac{e_- + p_+ }{e_+ + p_-} \, .
\ee

In the general case, the pressures and energy densities in the symmetric and broken 
phases are obtained from the free energy
 $\mathcal{F} = - p$ (see (\ref{pande})). 
In a concrete model (with known $\mathcal{F}$) the temperature $ T_N$ at 
which the phase transition happens can be determined using the standard
techniques\cite{fate1,fate2} (see section \ref{sec_4}), and the thermodynamic potentials 
can be calculated in the two phases. Still, there are three unknown quantities ($v_+$, $v_-$ and either $T_+$ or $T_-$)
and only two equations (\ref{eq:vvs0}), so
that up to this point all hydrodynamically viable solutions are
parametrized by one parameter, usually chosen to be the wall velocity $v_w$.

In order to obtain the behaviour of the system far from the wall, energy-momentum conservation in the plasma 
$\partial^\mu T_{\mu\nu}^{\mathrm{plasma}} = 0$ is imposed\footnote[3]{Away from the wall, $T_{\mu\nu}^{\phi}$
just gives a constant background that does not play any role in energy-momentum conservation.}, giving rise to a set of hydrodynamic
equations\cite{Landau,Steinhardt:1981ct,EKNS}. 
Since there is no intrinsic macroscopic length scale
present in the system, the solutions to the hydrodynamic equations are
self-similar, depending only on the combination $\xi = r/t$ (where $r$
denotes the radial coordinate of the bubble from its centre and $t$ the time since
nucleation), and the hydrodynamic equations read

\be
\label{eq:hydro}
\frac{1-\xi\, v(\xi)}{1-v^{2}(\xi)}\, \left[\frac{\mu^2(\xi,v)}{c_s^2}-1\right]\, \frac{\partial v}{\partial \xi} = 
2 \frac{v(\xi)}{\xi} \ , \quad \quad
\frac{4 \, \mu(\xi,v)}{1-v^{2}(\xi)}\, \frac{\partial v}{\partial \xi} = \frac{1}{w(\xi)}\,\frac{\partial w}{\partial \xi}
\, ,
\ee

\noindent where $c_s$ is the speed of sound of the plasma and $\mu$ is the Lorentz transformed plasma fluid velocity

\be
\label{eq:hydro2}
\mu(\xi,v) = \frac{\xi-v}{1-\xi\, v} \, .
\ee

\begin{figure}[thb]\begin{center}
\includegraphics[width=0.48\textwidth, clip]{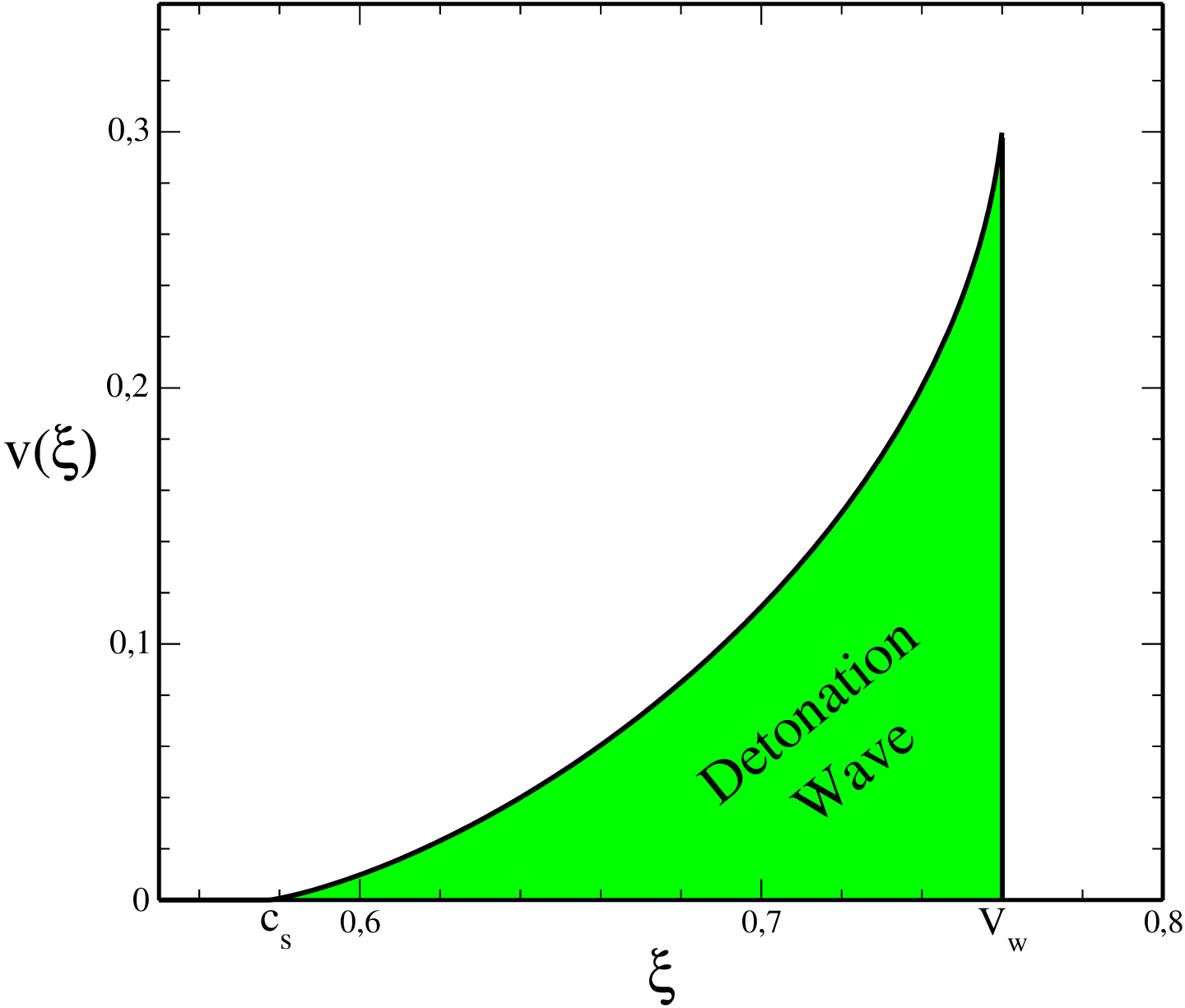} \hspace{4mm}
\includegraphics[width=0.48\textwidth, clip]{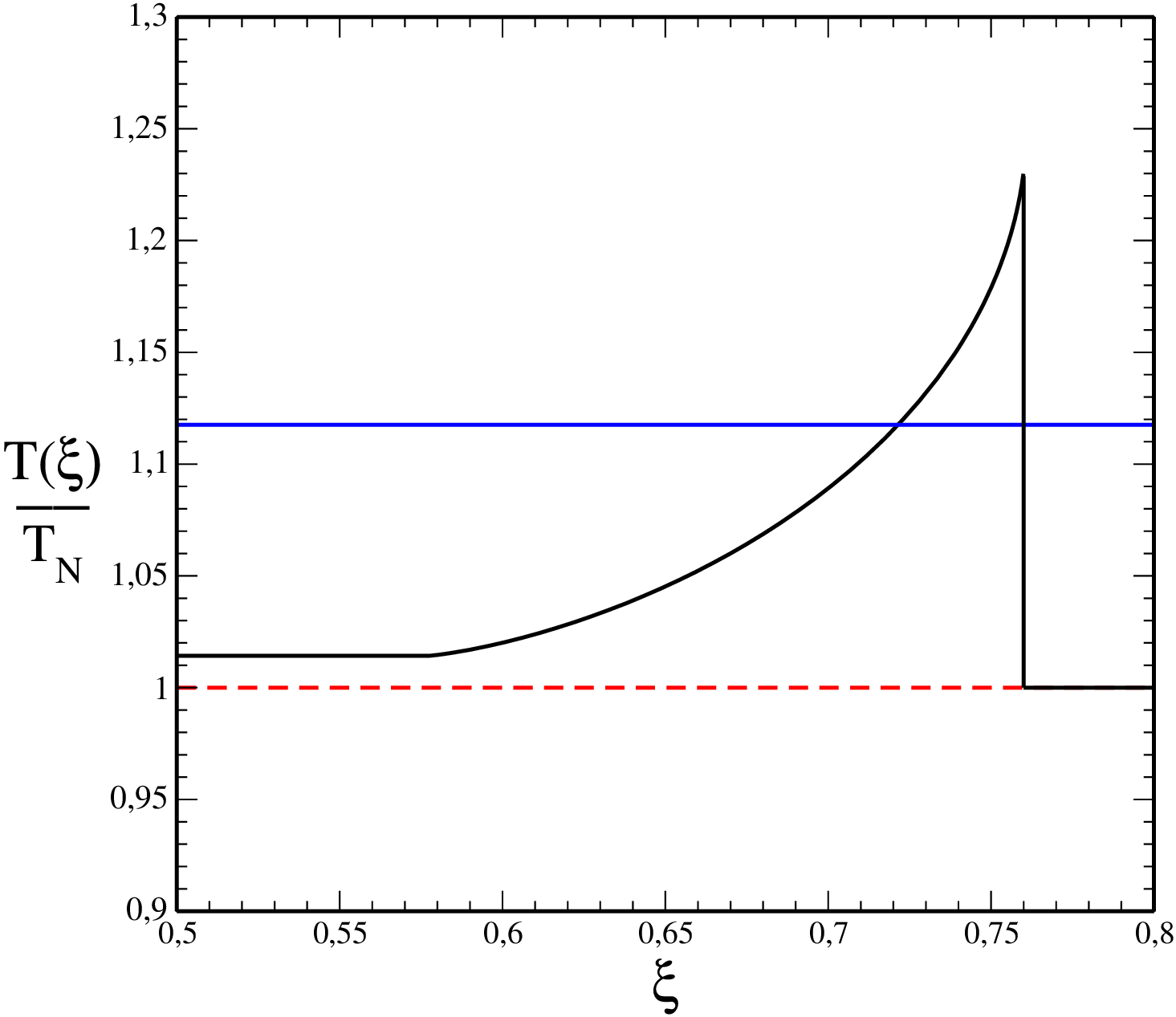}
\caption{\label{fig:1}
\small 
The solutions for the plasma velocity $v(\xi)$ (LEFT) and temperature $T(\xi)/T_N$ (RIGHT) in the case of a 
detonation (for $v_w = 0.76$ and $\alpha_N = 0.078$). The dashed-red horizontal line corresponds to $T_N$ and the solid-blue horizontal
line corresponds to $T_c$.}
\end{center}
\end{figure}

The solutions to (\ref{eq:hydro}) correspond to the velocity profile $v(\xi)$ and enthalpy profile 
$w(\xi)$ (or temperature profile $T(\xi)$) of the plasma away from the bubble wall.
Quite generally, there are three different types of
solutions (see \cite{EKNS} for details): detonations, deflagrations and hybrid
solutions. In detonations\cite{Steinhardt:1981ct}, the wall expands at supersonic 
velocities $v_w > c_s$ and the vacuum energy of the Higgs leads to a rarefaction wave behind
the bubble, while the plasma in front of the wall is at rest (see Figure \ref{fig:1}). In
this case, the wall velocity is $v_w = v_+ > v_- \geq c_s$ \cite{Laine:1993ey}, and the 
temperature outside the bubble wall corresponds 
to the temperature of the universe at the time of the phase transition $T_+ = T_N$. In
deflagrations\cite{Gyulassy:1983rq}, the plasma is mostly affected by reflection of
particles at the bubble wall and a compression wave builds up in front
of the wall, leading to $T_+ > T_N$. ``Pure" deflagrations are subsonic, with the 
plasma behind the wall at rest and $v_w = v_- > v_+$, while 
the hybrid case occurs for supersonic deflagrations\cite{KurkiSuonio:1995pp} where both effects (compression 
and rarefaction wave) are present.

\begin{figure}[thb]\begin{center}
\includegraphics[width=0.6\textwidth, clip]{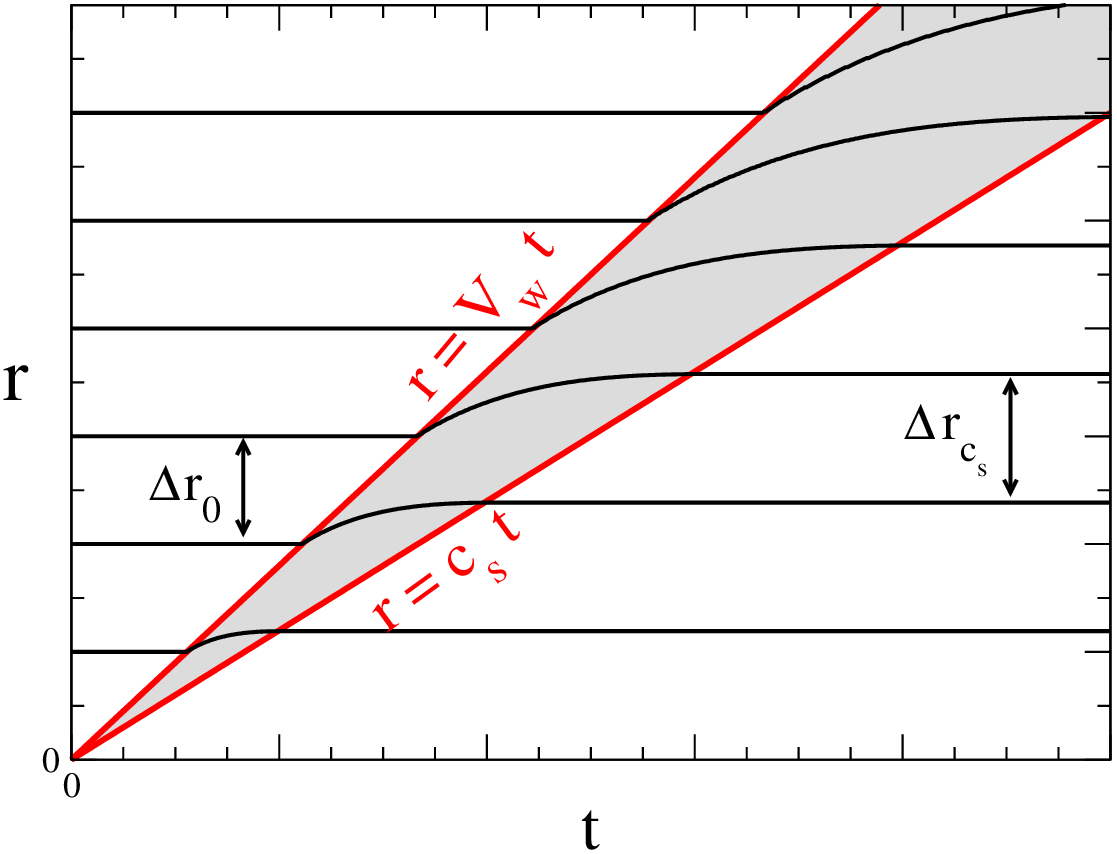}
\caption{\label{fig:2}
\small 
Motion of plasma volume elements (black lines) as the detonation wave (in grey) sweeps through them, for different values of 
the initial distance $r_0$ to the center of the bubble.}
\end{center}
\end{figure}

We are interested in the behaviour of a plasma volume element in the case of a detonation, 
since it will be used in the following of the analysis. In this case, a plasma volume element initially at 
rest at a distance $r_0$ from the center of the bubble will feel the 
arrival of the detonation wave at time $t = r_0/v_w$ and will be dragged by it, moving along the wave
according to 

\be
\label{eq:detonationelement1}
\frac{d\,r}{d\,t} = v(\xi = r/t) \, ,
\ee

\noindent and eventually coming back to rest (for $\xi = c_s$) at a distance $r > r_0$ from the center of the bubble, 
as shown in Figure \ref{fig:2} (the motion of plasma volume elements in the case of deflagrations 
and hybrids can be described in a similar way).

It is possible to extract 
the $\xi$ and $r_0$ dependence of a trajectory $r$ obtained from (\ref{eq:detonationelement1}), using the 
self-similarity of the detonation wave. From Figure \ref{fig:2}, we obtain the relation 

\be
\label{eq:hydro3}
\Delta r_{c_s} \equiv r(r_0 + \Delta r_0,\, \xi = c_s) - r  (r_0,\, \xi = c_s) = C({c_s}) \, \Delta r_0 \, ,
\ee 

\noindent where the constant $C({c_s}) >1$ does not depend on $r_0$. Due to the self-similarity, (\ref{eq:hydro3}) can be generalized to  
arbitrary values of $\xi$ inside the detonation wave

\be
\Delta r_{\xi} \equiv r(r_0 + \Delta r_0,\, \xi) - r  (r_0,\, \xi) = C (\xi) \, \Delta r_0 \, ,
\ee 

\noindent where the function $C(\xi)$ is again independent of $r_0$, and one has $C(\xi) > 1$ all along the detonation wave. It then 
follows that

\be
\label{eq:detonationelement2}
r (r_0, \xi) = C(\xi) \, r_0 + r (0,\xi) \simeq C(\xi) \, r_0 \, ,
\ee 

\noindent where $r (0,\xi)$ can be safely neglected\footnote[4]{If the 
distance of a plasma volume element to the center of the bubble is negligible compared to the 
final radius $R_B$ of the bubble ($r_0 \rightarrow 0$ for that volume element), then its distance to the 
bubble's center as it moves along the 
detonation wave is $r (0,\xi) \ll r_0$, generically for values of $r_0 \sim \mathcal{O}(10^{-4}R_B) - \mathcal{O}(R_B)$.}
 with respect to $r_0$. 
The result (\ref{eq:detonationelement2}) shows the explicit dependence of a trajectory $r(r_0,\,\xi)$ with the initial 
distance $r_0$, and will be used later in section \ref{sec_5}.

\section{Plasma Heating from Bubble Expansion}
\label{sec_3}

From Figure \ref{fig:1} it appears that the plasma temperature behind the 
wall (in the broken phase) is higher than the nucleation temperature $T_N$, as a consequence of the expansion of the bubble.
In this section we will perform a general analysis of the plasma heating behind the bubble wall, focusing on the case of 
bubbles expanding as detonations. Instead of performing the analysis in the general case $p = -\mathcal{F}$, 
we consider for simplicity the Bag EoS, holding if the plasma is a relativistic perfect fluid:

\be
\label{eosbag1}
p_+ = \frac{1}{3}a_+ T_+^4 - \epsilon  \quad  \quad e_+ = a_+ T_+^4 + \epsilon \, ,
\ee
\be
\label{eosbag2}
p_- = \frac{1}{3}a_- T_-^4 \quad \quad e_- = a_- T_-^4 \, ,
\ee

\noindent where $\epsilon = V_{0}(0)-V_{0}(\phi_0)$ denotes the false-vacuum energy resulting from the 
Higgs potential, and $a_\pm$ relate directly to the number of relativistic degrees of 
freedom $g^*$ in the symmetric and broken phases ($g_{\pm}^* = 30 \, a_{\pm} / \pi^2$). Using (\ref{eosbag1}) and (\ref{eosbag2}) 
we can write the matching equations (\ref{eq:vvs0}) as

\be
\label{eq:vvs}
v_+ v_- = \frac{1 - (1-3\alpha_+) b }
{3 - 3( 1 + \alpha_+ )b}  \quad \quad
\frac{v_+}{ v_-} = \frac{3  + (1-3\alpha_+) b}
{1 + 3(1 + \alpha_+ )b} \, ,
\ee 

\noindent where we defined 

\be
\alpha_+\equiv\frac{\epsilon}{a_+ T_+^4} \quad  \quad b\equiv\frac{a_+T_+^4}{a_-T_-^4} \, .
\ee

The quantity $\alpha$, being the rate of vacuum to thermal energy, 
typically characterizes the strength of the phase 
transition (the larger $\alpha$ the stronger the phase transition). Finally, the two equations 
in (\ref{eq:vvs}) can be combined to give

\be
\label{eq:vvs2}
v_+ = \frac{1}{1+\alpha_+}\left[ \left(\frac{v_-}{2}+\frac{1}{6
v_-}\right) \pm \sqrt{\left(\frac{v_-}{2}+\frac{1}{6 v_-}\right)^2 +
\alpha_+^2 +\frac{2}{3}\alpha_+ - \frac{1}{3}} \right]\, ,
\ee

\noindent so that there are two branches of solutions, corresponding to the $\pm$ signs in (\ref{eq:vvs2}). The presence
of a compression wave in front of the wall (deflagrations and hybrid solutions) corresponds to the $-$ branch
in (\ref{eq:vvs2}). In this case, the temperature just in front of the wall $T_+$ is higher
than the temperature of the universe $T_N$. Also, depending on the strength of the transition,
it is possible for the temperature just behind the wall $T_-$ to be larger than $T_N$. However, there 
is a hydrodynamic obstruction to having $T_- > T_c$\cite{KN}. This is clear since the 
average temperature in the wall is $\overline{T} \leq T_c$ 
(otherwise the net pressure on the wall that drives the expansion vanishes), and for a compression wave one 
has\footnote[5]{If $T_- > T_c$ would have been possible, 
this would have been potentially dangerous for EW baryogenesis scenarios due to the reactivation
of sphaleron processes in the broken phase behind the bubble wall.} 
$T_+ > \overline{T} > T_-$. 

For detonations the situation is quite different. In this case there is no compression wave in front of the wall, 
so $T_+ = T_N$ and $v_+ = v_w > v_- \geq c_s$ (with $c_s = 1/\sqrt{3}$ for a relativistic perfect fluid). 
The quantity $b = (a_+ T^4_N)/(a_- T^4_-)$, that measures the heating of the plasma behind the wall, is obtained from (\ref{eq:vvs})

\be
\label{rdeto}
b =  \frac{3 v_- v_w-1}{3 v_-v_w (1+\alpha_N)-(1-3 \alpha_N)} \, . 
\ee



\noindent If $\tilde{v}_w$ denotes the lowest possible detonation wall velocity, 
obtained for $v_{-} = c_s$, since $ \tilde{v}_w c_s < v_w v_- < 1$ it follows that 
$b $ is bounded by

\be
\frac{3 c_s\tilde{v}_w-1}{3 c_s\tilde{v}_w (1+\alpha_N)-(1-3 \alpha_N)} <b< \frac{1}{1 + 3 \alpha_{N}}  \, .
\ee

\noindent From the upper bound on $b$ it is clear that the plasma behind the wall is heated up 
with respect to the plasma in front, $T_- > T_N$ (since $a_+ > a_-$). We then want to know if for detonations it is possible
to achieve sufficient heating for $T_-$ to become larger than $T_c$. For the Bag EoS one has 

\be
\alpha_c = \frac{1}{3} \left(1- \frac{a_-}{a_+}\right) \, \longrightarrow \, 
T^4_c = T_N^4 \, \frac{3 \, \alpha_N}{1 - \frac{a_-}{a_+}} \, .
\ee
 
Then, using (\ref{rdeto}) the condition $T_- > T_c$ reads 

\be
\label{eq:Heating}
\frac{1}{\alpha_N} \left(\frac{a_+}{a_-} -1\right)\frac{3 v_-v_w (1+\alpha_N)-(1-3 \alpha_N)}{3 v_- v_w-1} > 1 \, .
\ee

\begin{figure}[thb]\begin{center}
\includegraphics[width=0.48\textwidth, clip]{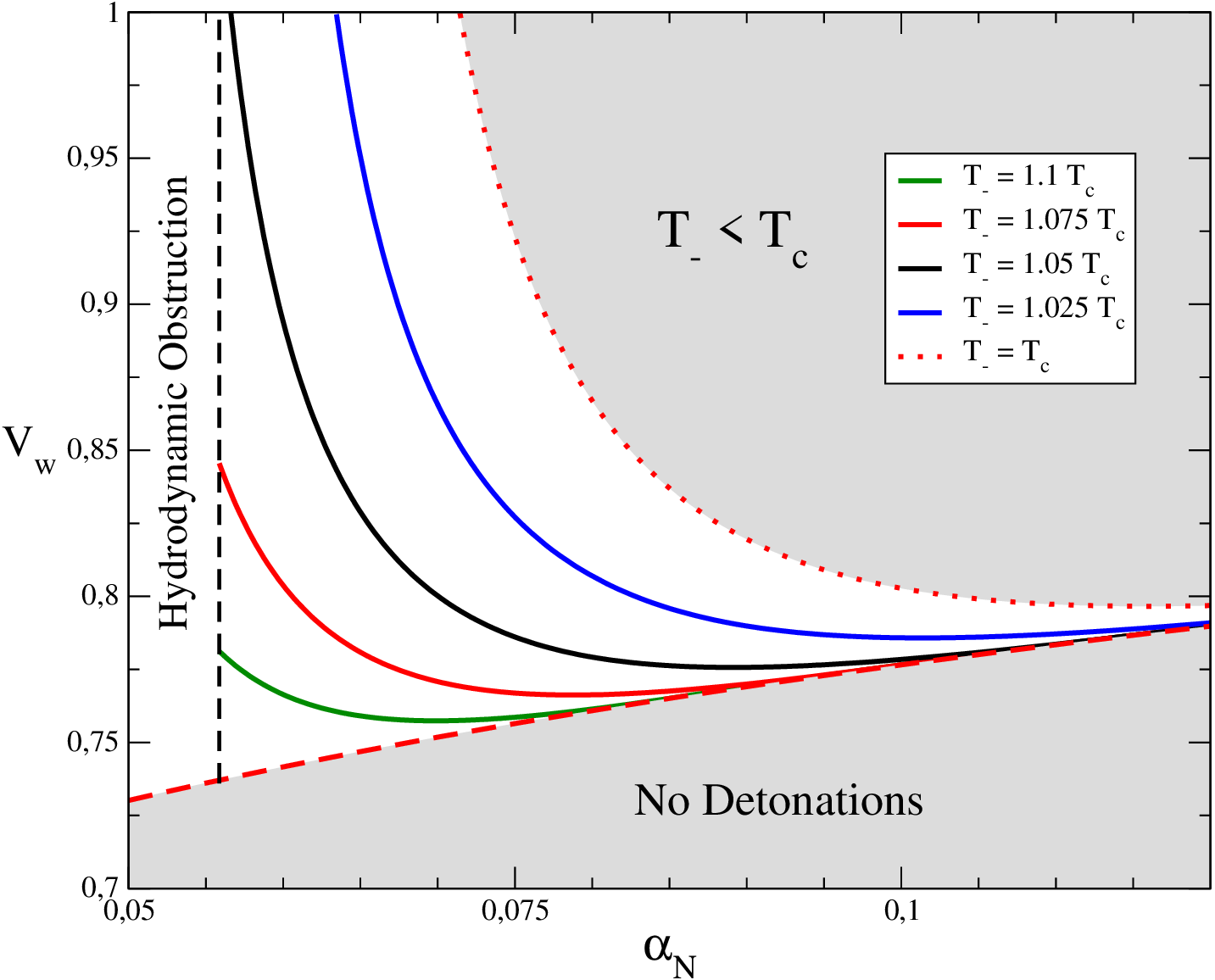}
\includegraphics[width=0.48\textwidth, clip]{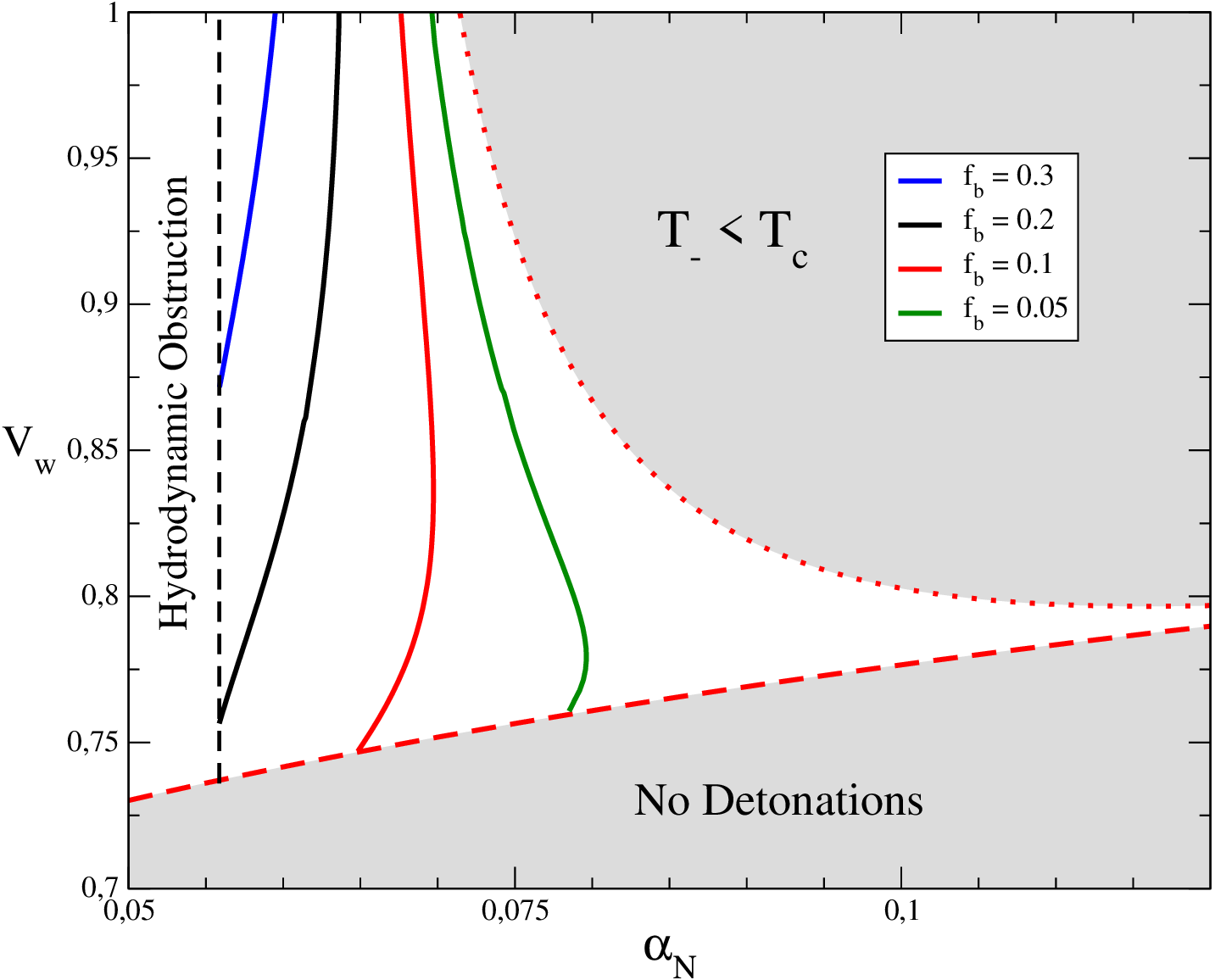}
\caption{\label{fig:3}
\small 
Region in the ($\alpha_N, v_w$)-plane for which $T_- > T_c$ (\ref{eq:Heating}), for $a_-/a_+ = 0.85$.
The dashed-red line corresponds to $v_w = \tilde{v}_w$ (lower bound on $v_w$), and the lower 
bound on $\alpha_N$ (dashed-black line) corresponds to the occurence of the hydrodynamic obstruction in the deflagration case\cite{KN}.
The different values of $f_b$ (RIGHT) correspond to different sizes of the 
region where $T > T_c$ with respect to the size of the bubble (\ref{cb}).} 
\end{center}
\end{figure}

\begin{figure}[thb]\begin{center}
\includegraphics[width=0.55\textwidth, clip]{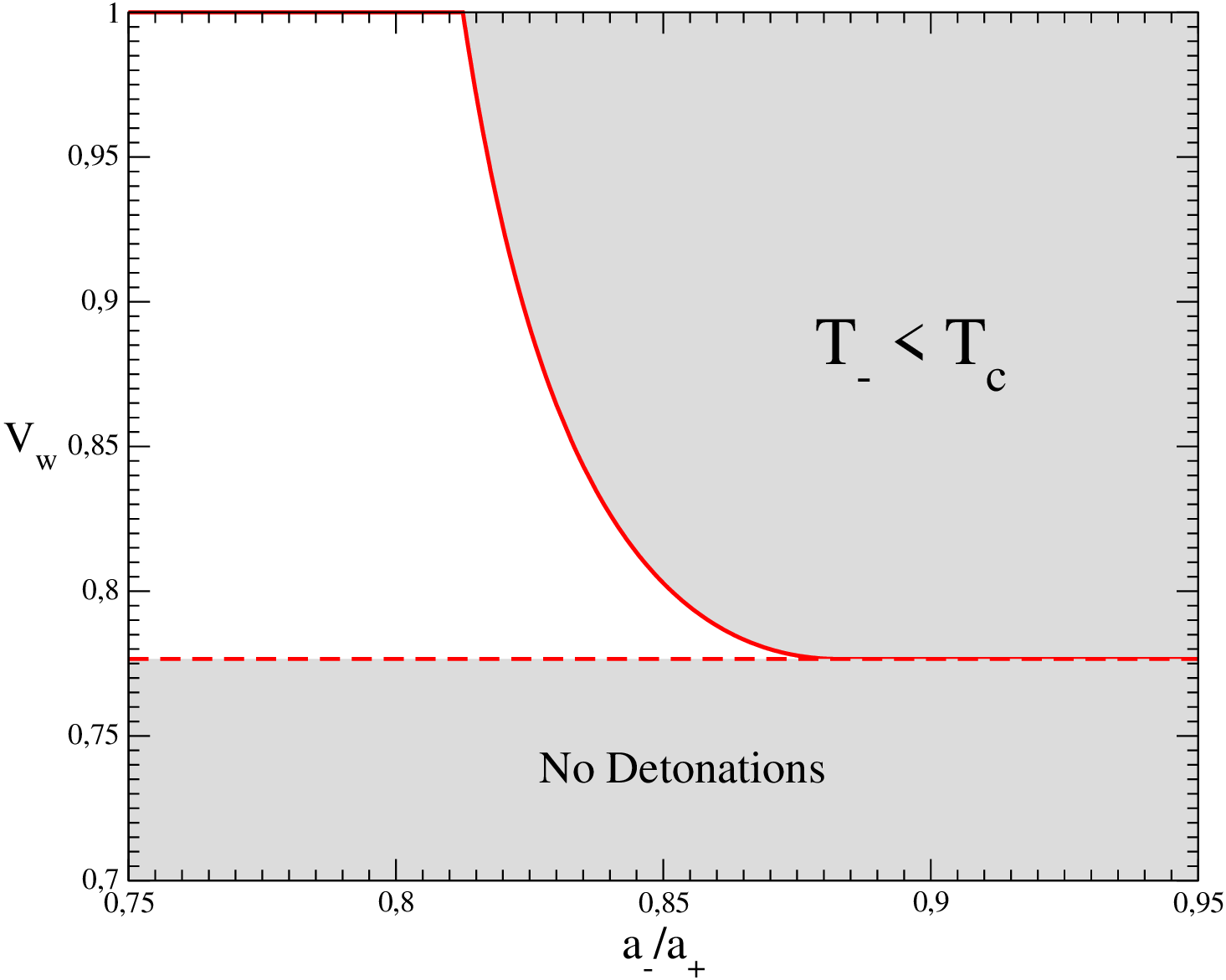}
\caption{\label{fig:aminus}
\small 
Region in the ($a_-/a_+, v_w$)-plane for which $T_- > T_c$ (\ref{eq:Heating}), for $\alpha_N = 0.1$.
As in Figure \ref{fig:3}, the dashed-red line corresponds to $v_w = \tilde{v}_w$ (lower bound on $v_w$).} 
\end{center}
\end{figure}

The region of the ($\alpha_N, v_w$)-plane (for a fixed ratio $a_-/a_+ = 0.85$) 
for which $T_- > T_c$ is verified is shown in Figure \ref{fig:3}. Apart from
the trivial bound $\alpha_N > \alpha_c$ ($T_N < T_c$ is needed for the bubbles to have been nucleated) 
there is a lower bound on $\alpha_N$ (dashed-black line in Figure \ref{fig:3}) due to the hydrodynamic obstruction 
that occurs in the deflagration case\footnote[6]{For $\alpha_N$ below this bound, the bubble will always expand as a subsonic deflagration,
no matter how small the friction on the wall (due to the plasma) is. Therefore, a detonation is never realized in this case \cite{KN}.}. 
We now consider the temperature profile of the plasma $T(\xi)$ in the rarefaction wave, evolving from  $T_- = T(v_w)$ down to
$T(c_s) < T_c$. In the cases when $T_- > T_c$, the condition $T(\xi_c) = T_c$ will be satisfied by some $\xi_c$ in the interval
$c_s < \xi_c < v_w$, and the size of the region $T(\xi) \geq T_c$ with respect to the size of the bubble
will simply be

\be
\label{cb}
f_b = \frac{v_w - \xi_c}{v_w} \, .
\ee
From Figure \ref{fig:3} it appears that the region where the heating can be sufficient to 
drive $T_{-}$ above $T_c$ becomes smaller as the strength of the phase 
transition grows ({\it i.e.} for increasing $\alpha_N$). This counterintuitive effect is 
due to the interplay between $\alpha_N$ and $a_-/a_+$. As shown in Figure \ref{fig:aminus}, for fixed $\alpha_N$ the region in the 
($a_-/a_+, v_w$)-plane where $T_-$ overcomes $T_c$  becomes larger for decreasing $a_-/a_+$. This behaviour is general and 
occurs also for stronger phase transitions. In a realistic model, 
$\alpha_N$ and $a_-/a_+$ will not be independent quantities, and these two opposite effects will combine, either to reduce 
the heating effect for stronger phase transitions, or to enhance it depending on the details of the model (see section \ref{sec_7}). 

Having established that for detonations it is possible to heat the plasma behind the wall above $T_c$, 
we now proceed to study the consequences of this heating 
effect: in particular, we investigate whether the system can go back to the symmetric phase locally in the 
regions where $T(\xi) \geq T_c$. 

\section{Nucleation of Symmetric Bubbles.}
\label{sec_4}

\subsection{Bubble Nucleation in the Standard Case.}
\label{symmetric_nucleation1}

\begin{figure}[thb]
\begin{center}
\includegraphics[width=8.cm,height=6.cm, clip]{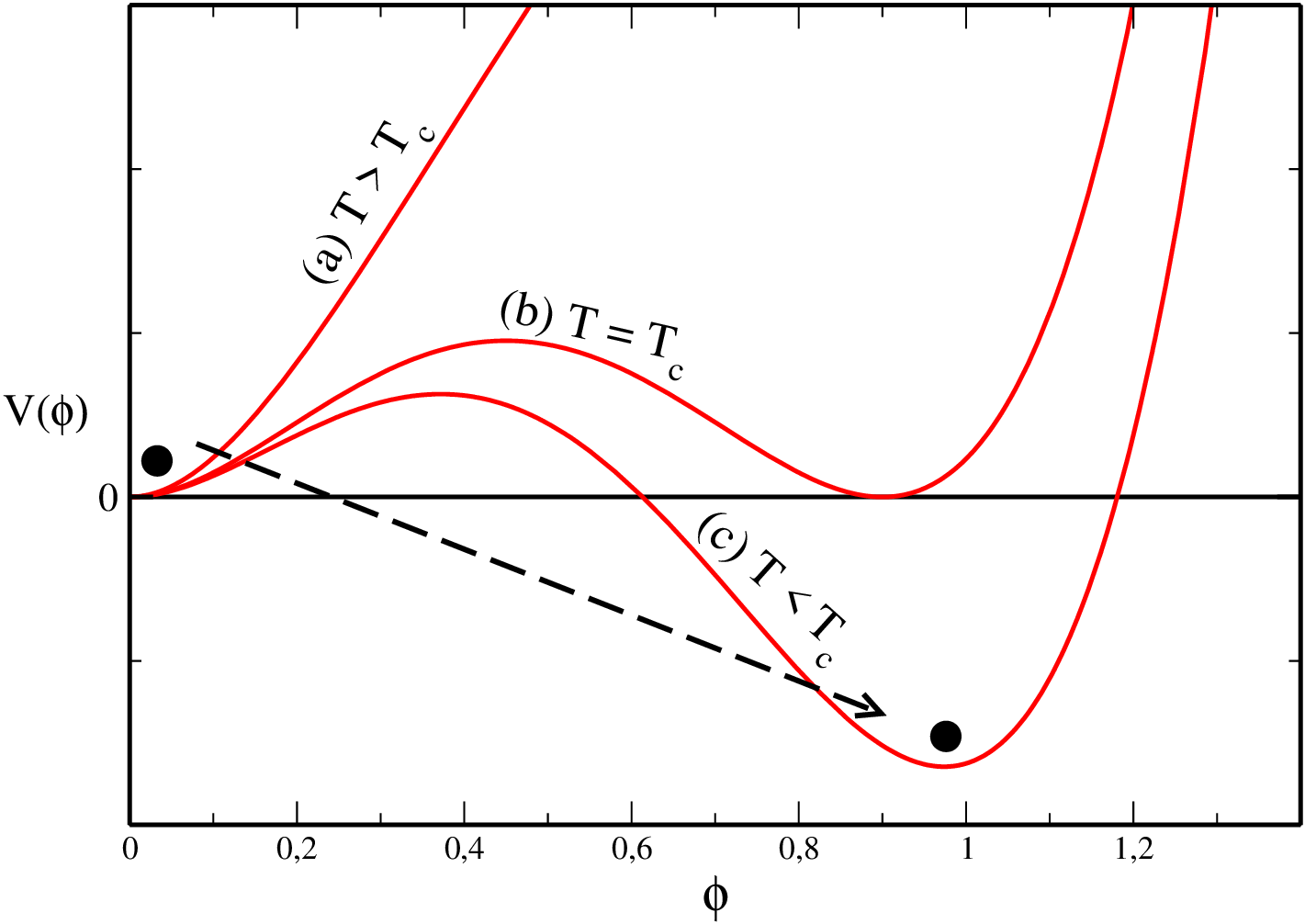}\hspace{3mm}
\includegraphics[width=8.cm,height=6.cm, clip]{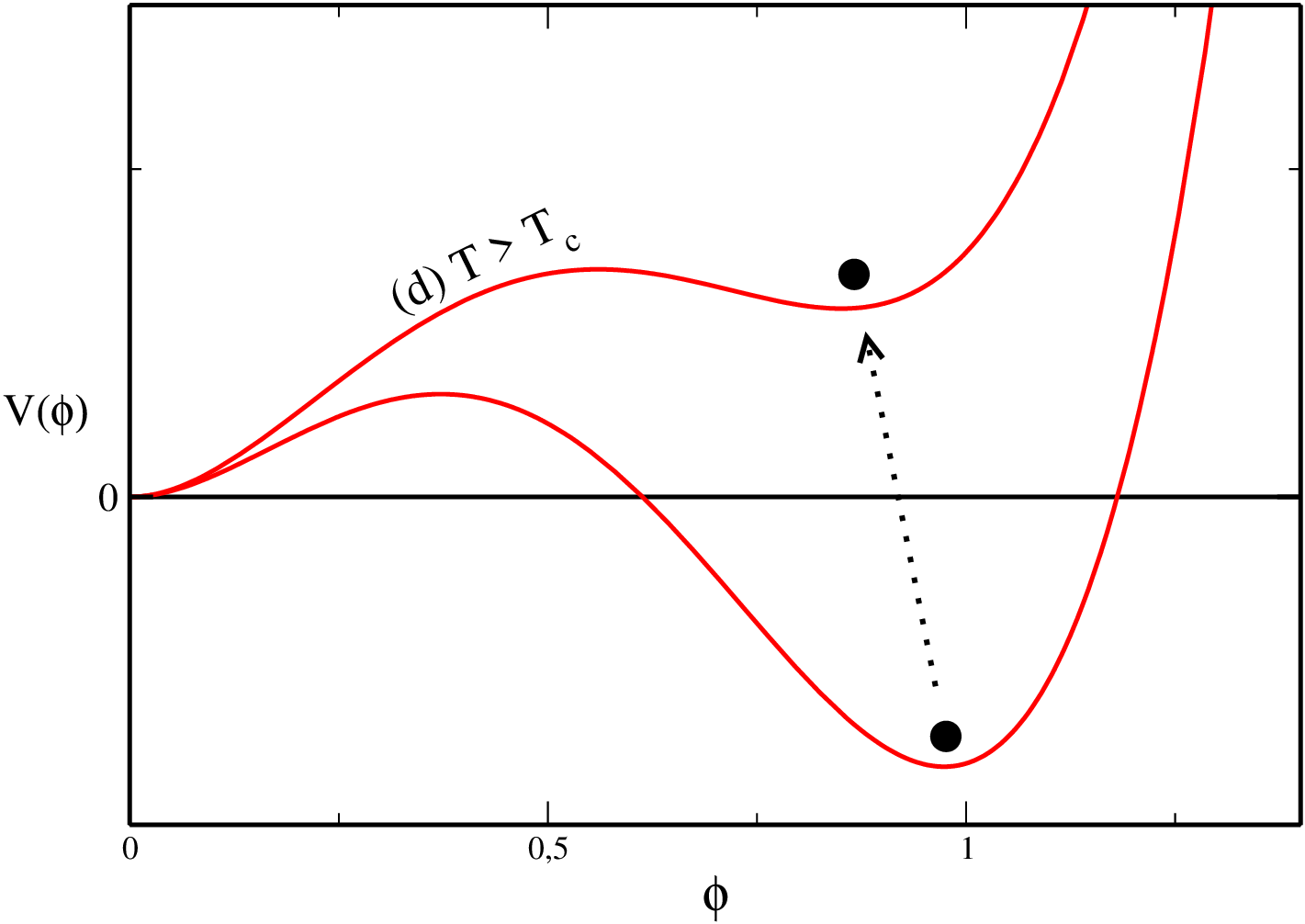}
\caption{\label{fig:4}
\small
LEFT: Evolution of the potential $V(\phi,T)$ with temperature in the case of a first
order phase transition in the early universe (see section \ref{symmetric_nucleation1}). 
RIGHT: Local change in the potential for the case $T_{-} > T_c$ (see section \ref{symmetric_nucleation2}).}
\end{center}
\end{figure}

Consider the behaviour with temperature of a potential $V(\phi,T)$ giving rise to a first order phase transition
in the early universe (being $\phi$ the order parameter of the transition), such as the one depicted in Figure \ref{fig:4}. 
For very high temperatures the only minimum is at $\phi = 0$ (Figure \ref{fig:4} (a)). As the universe expands and the temperature lowers,
a new minimum $\phi_B$ develops away from the origin and eventually the system reaches a critical temperature $T_c$ for which
$V(0) = V(\phi_B)$ (Figure \ref{fig:4} (b)). Then, for $T < T_{c}$  
the potential has a metastable minimum at $\phi = 0 $ and a stable one at $\phi \neq 0 $, 
separated by a potential barrier (Figure \ref{fig:4} (c)) and it becomes possible to 
tunnel from $\phi = 0$ to the stable minimum at $\phi_B$.   
The nucleation probability per unit time and volume from the metastable 
minimum to the stable one is given at finite temperature by \cite{fate2}

\be
\label{eq:nucleation}
\Gamma(T) \simeq T^4 \, \left( \frac{S_{3}(T)}{2 \, \pi \,T} \right)^{3/2} \textrm{exp}\left(-\frac{S_{3}(T)}{T} \right) \, ,
\ee

\noindent with $S_{3}(T) $ being the tunneling three dimensional Euclidean action\cite{fate1,fate2}. Then, in the radiation dominated era 
the mean number of true (broken) vacuum bubbles $N_{\textrm{Bubbles}}$ 
nucleated in a volume $V $ during a time interval $t-t_{c}$ (being $t_{c}$ the time at which $T = T_{c}$) would be simply 

\be
\label{eq:BubbleNumber}
N_{\textrm{Bubbles}} = \int_{t_{c}}^{t} \Gamma(\overline{t})\, V(\overline{t}) \,d\overline{t} = \int_{T}^{T_{c}} 
\frac{H^{-1}(\overline{T})}{\overline{T}} \, V(\overline{T})
 \, \Gamma(\overline{T}) \, d\overline{T} \, ,
\ee

\noindent where the Hubble rate $H(T)$ and the time-temperature relation during the radiation dominated era read

\be
H(T) = \sqrt{\frac{32 \, \pi^3}{90}} \frac{\sqrt{g_+ \, T^4}}{M_{pl}} \quad \quad \quad \quad t = \frac{1}{2 \, H(T)} \, .
\ee

Even though nucleation of broken phase bubbles becomes possible as soon as the 
temperature drops below the critical temperature $T_{c} $, the actual nucleation 
temperature $T_{N} $ at which the phase transition starts is defined 
as the one for which the mean number of bubbles nucleated inside a Hubble volume $H^{-3}(T)$
is of $\mathcal{O} (1)$ \cite{AndersonHall}

\be
\label{eq:TN1}
N_{\textrm{Bubbles}}(T_{N}) = \int_{T_{N}}^{T_{c}} \frac{T^3}{H^4(T)} \, 
\left( \frac{S_{3}(T)}{2 \, \pi \,T} \right)^{3/2} \textrm{exp}\left(-\frac{S_{3}(T)}{T} \right) dT \simeq 1 \, ,
\ee

\noindent and this in turn leads to the condition

\be
\label{eq:TN2}
\frac{S_{3}(T_N)}{T_N} \simeq 142 \, .
\ee
%

\subsection{Symmetric Bubble Nucleation.}
\label{symmetric_nucleation2}

Consider now one of these nucleated bubbles of broken phase 
expanding in the plasma (from now on we will refer to these bubbles as ``broken" bubbles). 
Recalling section 3, if the bubble expands as a detonation the plasma behind the 
bubble wall gets heated up, and if $T_{-} > T_c$ (being $T_{-} = T(v_w)$ the temperature just behind the bubble wall) 
then for some region $\xi_c < \xi < v_w$ of the detonation wave one has 
$T(\xi) > T_c$ (with $T(\xi_c) = T_{c}$). 
In this region the system is then in situation (d) of Figure \ref{fig:4} (RIGHT), and can tunnel 
back to the symmetric minimum, with the subsequent formation of
bubbles of the symmetric phase inside the broken bubble
(from now on we will refer to these bubbles as ``symmetric" bubbles). 
Differently from the case discussed in the previous section,  
the available volume for the inverse tunneling and subsequent formation of the 
symmetric bubbles is not the entire Hubble volume but just the region inside the  
broken bubble where $T(\xi) > T_c$, whose volume is $V_{f_b}$. Recalling (\ref{cb}) and the fact that 
at a given time $t$ after the nucleation of the broken bubble its radius 
will be $R_B \equiv v_w \, t$, (and we also define $R_{\xi_c}\equiv \xi_c \, t$),
$V_{f_b}$ can be written as 

\be
V_{f_b} =  \frac{4 \, \pi}{3} R_B^3 -  \frac{4 \, \pi}{3} R_{\xi_c}^3 
 = \frac{4 \, \pi}{3} t^3 \left(  v_w^3 - \xi_c^3  \right)
\simeq 4 \, \pi\, (v_w\, t)^3  \, f_b\, \left( 1-f_b\right)\, .
\ee

The nucleation probability (\ref{eq:nucleation}) is not constant over $V_{f_b}$ (due to the 
temperature variation along the detonation wave), and so 
the mean number of nucleated symmetric 
bubbles inside a broken bubble is not defined by (\ref{eq:BubbleNumber}). 
Instead, the number of symmetric bubbles nucleated inside the broken bubble {\it per unit time} 
is obtained by integrating (\ref{eq:nucleation})
over the region of volume $V_{f_b}$

\be
\label{eq:nucleation2}
N^S_{\textrm{B}}(t) = 
4 \,\pi \int_{R_{\xi_c}}^{R_B} \Gamma\left( T(R)\right) \,R^2 \,dR \, ,
\ee

\noindent which then gives

\be
\label{eq:nucleation3}
N^S_{\textrm{B}}(t) = 4 \,\pi \,t^3
\int_{\xi_c}^{v_w} \Gamma\left( T(\xi)\right)\, \xi^2 \,d\xi = 4 \,\pi \,t^3 \, 
\int_{T_c}^{T_-} \Gamma\left( T\right) \, \frac{d\xi}{dT} \, \xi(T)^2 \, dT \, .
\ee

The total number of symmetric bubbles nucleated inside the broken bubble up to a time $\tau$ (since the nucleation of the 
broken bubble\footnote[7]{Actually, since symmetric bubbles cannot nucleate until the broken bubble reaches a stationary state
and the detonation wave is established, $\tau$ is really the time since the broken bubble has reached a stationary state, and not the 
time since nucleation. However, both are the same for all practical purposes, since the stationary state 
is reached extremely fast compared to the duration of the phase transition.}) $\mathcal{N}^{S}_{\textrm{Bubbles}}(\tau)$ 
is just found by integrating (\ref{eq:nucleation3}) over time

\be
\label{eq:nucleation4}
\mathcal{N}^{S}_{\textrm{Bubbles}}(\tau) = \int_{0}^{\tau} dt \, N^S_{\textrm{B}}(t) \, .
\ee

Recalling (\ref{eq:TN1}) and (\ref{eq:TN2}), (\ref{eq:nucleation4}) can be approximately rewritten as

\be
\label{eq:nucleation5}
\mathcal{N}^{S}_{\textrm{Bubbles}}(\tau) \simeq 
\frac{\tau^4}{H^{-4}} \,
\textrm{exp}\left(142 - \frac{S^S_{3}(T_{-})}{T_{-}}\right) \int_{T_c}^{T_-} \frac{\Gamma\left( T\right)}{\Gamma\left(T_{-}\right)}
\,\frac{d\xi}{dT} \, \xi(T)^2 \,dT \, ,
\ee

\noindent where $S^S_{3}(T)$ is the tunneling action from the broken to the symmetric phase. Since $\tau \leq \beta^{-1}$, 
where $\beta^{-1}$ is the duration of the phase transition 
($\beta/H\sim 10 - 1000$ \cite{Hogan:nucleation}), then from (\ref{eq:nucleation5}) 
the value of $S^S_{3}(T_{-})/T_{-}$
needed to nucleate at least one symmetric bubble per broken bubble is bounded by 

\be
\label{eq:nucleation7}
\frac{S^S_{3}(T_{-})}{T_{-}} \lesssim 142 - 4 \, \mathrm{log} \left(\frac{\beta}{H}\right) \sim 125 \, ,
\ee
where we have neglected the contribution from the integral in (\ref{eq:nucleation5}): this 
would in principle further suppress the above value for $S^S_{3}(T_{-})/T_{-}$ due to the fact 
that the integrand is strongly peaked at $T\rightarrow T_-$, but this would enter in (\ref{eq:nucleation7}) only 
logarithmically. The above equation shows that, due to the volume suppression, it is generically more difficult 
to tunnel to the symmetric phase. 

A symmetric bubble can in principle be nucleated anywhere in the volume $V_{f_b}$. However,  
the situation is different from the case of broken bubble nucleation, which is a true random 
process since the nucleation probability is the same everywhere. Here, the nucleation probability is much higher 
close to $\xi = v_w$ than close to $\xi = \xi_c$, because one has $T(v_w) > T(\xi_c)$ and the nucleation probability $\Gamma(T)$
varies exponentially with $T$ (\ref{eq:nucleation}). The average nucleation
value for $\xi$ is then very close to $v_w$

\be
\label{eq:nucleation6}
\frac{\int_{\xi_c}^{v_w} \Gamma\left( T(\xi)\right) \, \xi \, d\xi }{\int_{\xi_c}^{v_w} \Gamma\left( T(\xi)\right) \, d\xi}
\simeq v_w \, .
\ee
Most of the symmetric bubbles are then nucleated very close to the broken bubble wall.

\section{Growth of Symmetric Bubbles.}
\label{sec_5}

Let us assume that a symmetric bubble nucleates at $\xi_S \simeq v_w$ (at a given time $t^S_{nucl}$ 
after the nucleation of the broken bubble) and subsequently expands. Due to the inhomogeneity and anisotropy of the background in 
which the symmetric bubbles grow, the expansion will render them non-spherical. 
As a consequence, the symmetric bubble expansion cannot be described by a self-similar
solution in a way analogous to the case of the broken bubble (recall section \ref{sec_2}). In fact, a rigorous solution of the system 
including the symmetric bubble expanding inside the broken bubble would require 
solving $\partial_{\mu} T^{\mu\nu} = 0$ over the whole system (with boundary conditions on the symmetric and broken bubble
walls), a very difficult problem both analytically and numerically. We expect however that a good approximation is to consider 
the broken bubble as a background on which the symmetric bubble expands, neglecting the back-reaction on this background
due to the symmetric bubble expansion. This is certainly the case in regions of the detonation wave of the 
broken bubble far from the symmetric bubble. Moreover, the symmetric 
bubble expansion turns out to be very slow (as we will see later in this section), and 
so it is reasonable to expect that the perturbation of the 
background 
is small. This is indeed similar to the case of broken bubbles 
expanding as very slow deflagrations, where the perturbation of the temperature background 
due to the bubble expansion is found to be very small \cite{EKNS} (not surprisingly, since the perturbation has to vanish
in the limit $v_w \rightarrow 0$). 
In the following we therefore consider that back-reaction is small and we neglect it, since this is justified even when a large amount of symmetric bubbles are nucleated\footnote[8]{We also disregard here the perturbation on the background due to possible collisions between the symmetric bubbles. 
These collisions are essentially absent when few symmetric bubbles are nucleated inside each broken
bubble, but some collisions may occur if the number of symmetric bubbles is very large.}.

Even though for $T > T_c$ the free energy of the system in the symmetric phase $\mathcal{F}(\phi=0,T)$ is
smaller than the one in the broken phase $\mathcal{F}(\phi_B,T)$ (and inverse tunneling does indeed occur),
the latent heat of the inverse transition is negative, due to the fact that for the symmetric
bubble the broken and symmetric phases are interchanged (see Appendix A for details). Then, it is clear
that the expansion of the symmetric bubbles 
cannot be fueled by the liberation of latent heat as the bubble expands (this is analogous to an endothermic reaction
in the context of slow combustion \cite{Landau}), and in this case one 
might be doubtful that the expansion of the symmetric bubbles actually takes place. 

However, there are two facts that question this last conclusion. First, there is an increase in the thermal 
energy of the system as the symmetric bubbles expand (due to the increase in the number of relativistic degrees of freedom of 
the plasma when it gets converted to the symmetric phase), which contributes to the pressure difference on the 
symmetric bubble wall driving the expansion. Second, since the heating of the plasma due to the expansion of the 
broken bubble keeps the temperature of the background in which the symmetric bubbles expand above $T_c$ for 
$\xi_c < \xi < v_w$, the temperature of the plasma in this region is effectively maintained 
despite the endothermic nature (negative latent heat) of the symmetric bubble expansion, making the expansion possible.
In this scenario the expansion velocity of the symmetric bubble
with respect to its background can be estimated qualitatively to be

\be
\label{eq:growth1}
\delta v(\xi) = \frac{T(\xi)}{T_c} -1\, .
\ee

This can be understood as follows (as argued in \cite{Witten:1984rs}): as the symmetric bubble expands, 
the endothermic nature of the expansion
will cause the temperature close to the symmetric bubble wall to drop momentarily. If the temperature would drop down to $T_c$, 
this would prevent further expansion \cite{KN}. However, the background temperature $T(\xi)$ close to the wall gets re-established 
by heat transport (since the expansion of the broken bubble keeps the background constant) 
and the symmetric bubble expansion continues. Since the rate of heat transport is roughly proportional 
to $(T(\xi)-T_{c})/T_{c}$, the velocity of symmetric bubble growth
will then be of the order of $(T(\xi)-T_{c})/T_{c}$ if the temperature difference is 
small\footnote[9]{Note that the situation described here is reversed from the one of \cite{Witten:1984rs}, which deals with  
the expansion of usual broken bubbles and analyses the case of ambient 
temperature $T < T_c$. In that case, the liberation of the (positive) latent heat causes 
the plasma right in front of the wall to be heated up to $T_c$, momentarily 
preventing further expansion. The liberated heat has then to be carried away into the symmetric phase
for the expansion to continue.}\cite{Witten:1984rs,Landau,KN}.

\begin{figure}[ht]
\begin{center}
\includegraphics[width=0.75\textwidth, clip ]{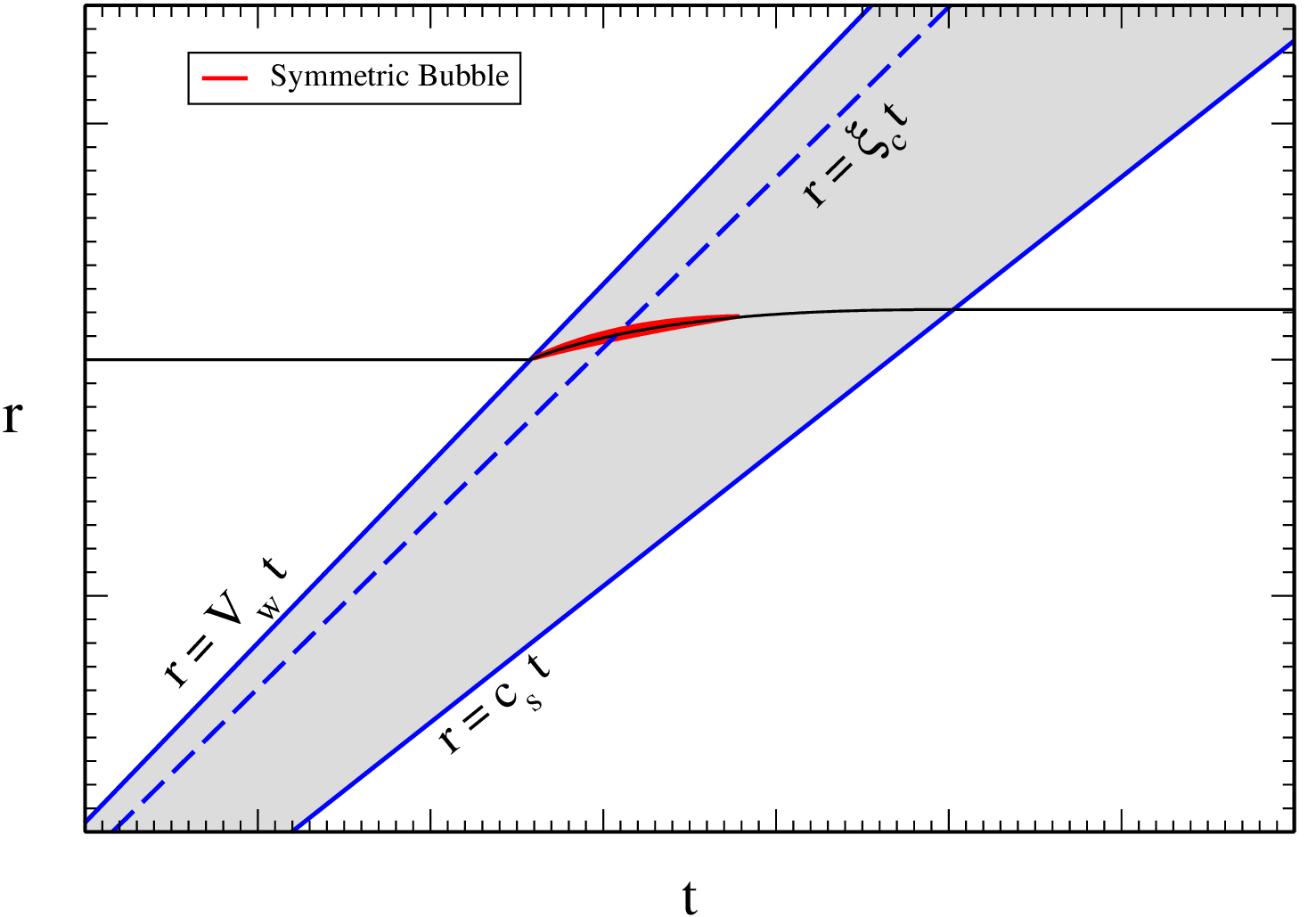}
\caption{\label{fig:5}
\small Detonation wave of the broken bubble (in grey), and the symmetric bubble nucleated inside (in red).
The solid-black line corresponds to the evolution of $r_{c}$ (\ref{eq:growth2}) (for $v_w = 0.76$ and $\alpha_N = 0.078$).}
\end{center}
\end{figure}

\begin{figure}[ht]
\begin{center}
\includegraphics[width=0.4\textwidth, clip ]{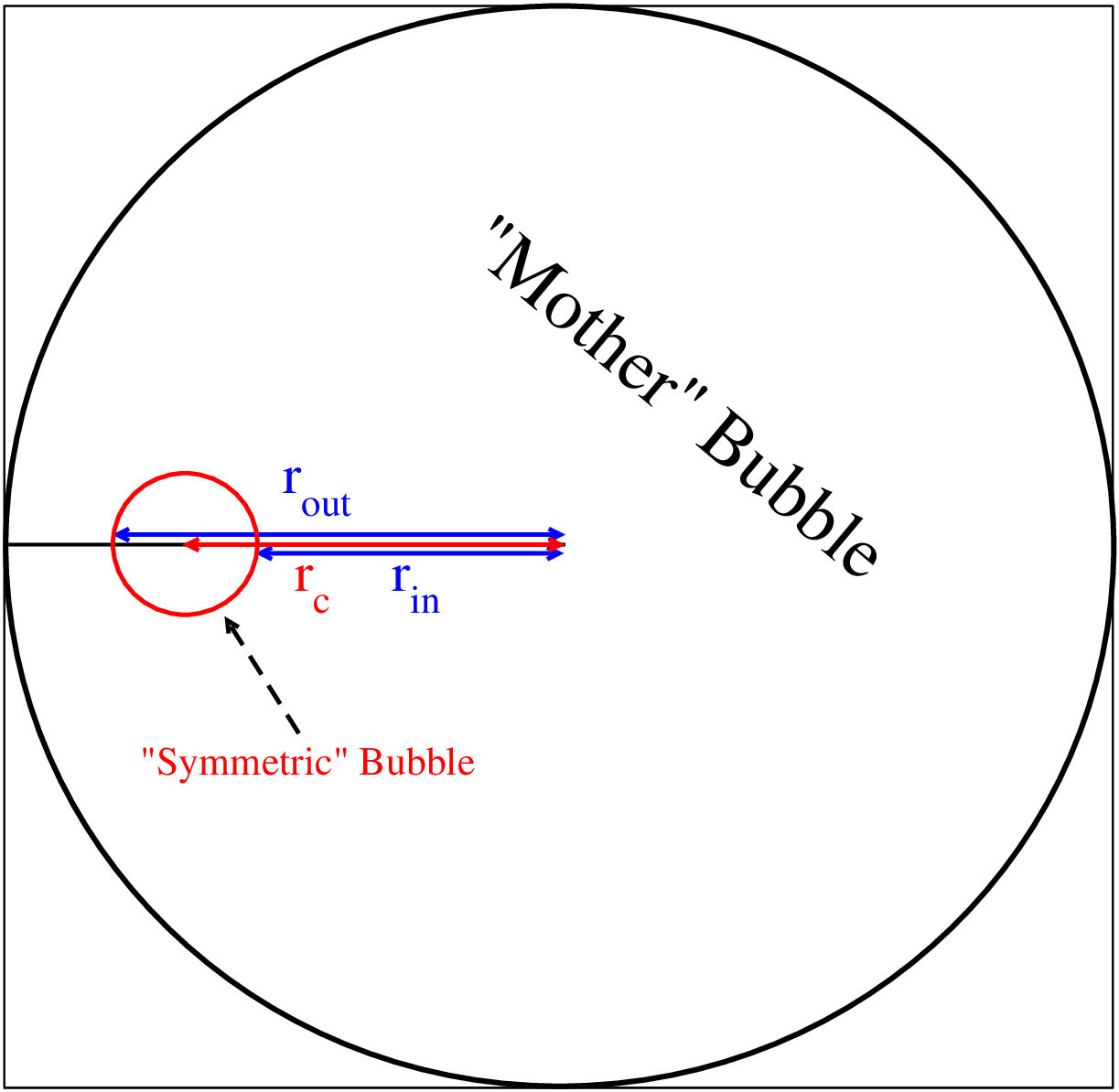}
\caption{\label{fig:6}
\small Sketch of the expansion of a symmetric bubble inside a broken bubble, and definition of 
$r_{c}$, $r_{\mathrm{in}}$ and $r_{\mathrm{out}}$.}
\end{center}
\end{figure}

\begin{figure}[ht]
\begin{center}
\includegraphics[width=0.8\textwidth, clip ]{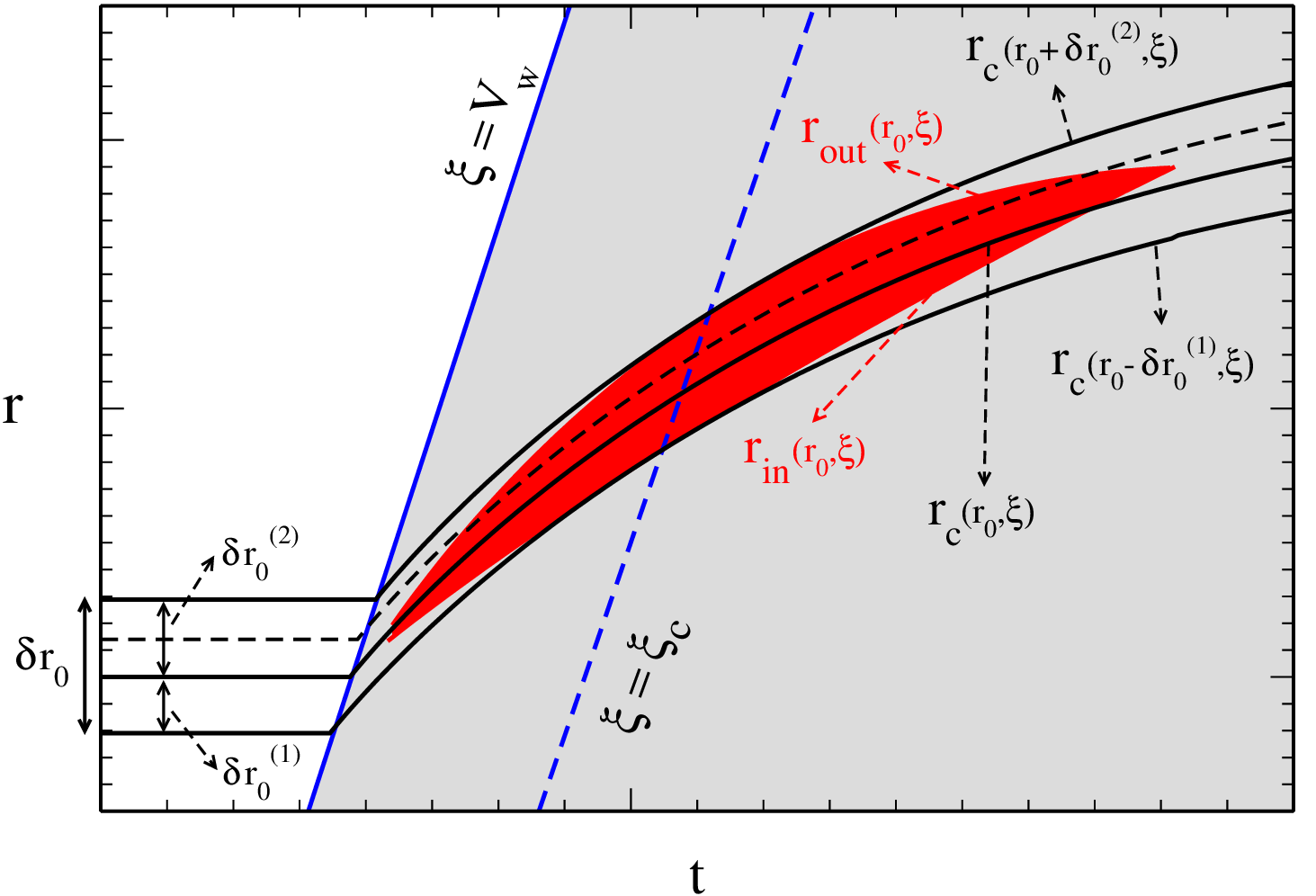}
\caption{\label{fig:7}
\small
Detail of Figure \ref{fig:5}. The symmetric bubble is depicted in red, its upper boundary corresponding to the evolution 
of $r_{\mathrm{out}}$ (\ref{eq:growth3}) and its lower boundary corresponding to the evolution of $r_{\mathrm{in}}$ (\ref{eq:growth3}).
Plasma volume elements with trajectories $r_c(\tilde{r}_0,\xi)$ contained between the upper and lower solid-black lines go through the
symmetric bubble (dashed-black line).}
\end{center}
\end{figure}

Symmetric bubbles nucleate at rest with respect to the plasma in which nucleation occurs, and move along with it.
The evolution of a symmetric bubble is shown in Figures \ref{fig:5} and \ref{fig:7}. 
Recalling (\ref{eq:detonationelement1}), the plasma element in which the symmetric bubble will be nucleated 
moves along the detonation wave of the broken bubble according to

\be
\label{eq:growth2}
\frac{d\,r_{c}}{d\,t} = v(\xi = r_{c}/t) \, ,
\ee

\noindent where $v(\xi)$ is the velocity profile of the detonation wave and $r_c(r_0,\xi)$ is the distance between the plasma element 
and the center of the broken bubble\footnote[10]{It is also effectively the distance between the center of the 
symmetric bubble and the center of the broken bubble (see Figure \ref{fig:6})} 
(being $r_0$ its initial value, when the plasma element is at rest before the arrival
of the detonation wave - see section \ref{sec_2}). The plasma volume element in which the symmetric bubble is nucleated 
passes through the detonation wave following a trajectory like those depicted in Figure \ref{fig:2} (solid-black lines), 
moving initially through the region where $T(\xi) > T_c$ and entering 
the region with $T(\xi) < T_c$ at a later time (see Figure \ref{fig:5}). 

Then, focusing on the expansion of the symmetric bubble along the broken bubble's radial direction
(see Figure \ref{fig:6}), 
the point of the symmetric bubble wall closest to the broken bubble center
$r_{\mathrm{in}}$ and the point of the symmetric bubble wall closest to the broken bubble wall $r_{\mathrm{out}}$ evolve as

\be
\label{eq:growth3}
\frac{d\,r_{\mathrm{in}}}{d\,t} = 
\frac{v\left(\frac{r_{\mathrm{in}}}{t}\right) - \delta v\left(\frac{r_{\mathrm{in}}}{t}\right)}
{1- v\left(\frac{r_{\mathrm{in}}}{t}\right)\, \delta v\left(\frac{r_{\mathrm{in}}}{t}\right)}
\quad \quad \quad \quad
\frac{d\,r_{\mathrm{out}}}{d\,t} = 
\frac{v\left(\frac{r_{\mathrm{out}}}{t}\right) + \delta v\left(\frac{r_{\mathrm{out}}}{t}\right)}
{1+ v\left(\frac{r_{\mathrm{out}}}{t}\right)\, \delta v\left(\frac{r_{\mathrm{out}}}{t}\right)}
\, ,
\ee

\noindent with $\delta v$ given by (\ref{eq:growth1}). As seen from Figure \ref{fig:7}, while the 
symmetric bubble is in the region $\xi > \xi_c$, where $T(\xi) > T_c$, 
the plasma flows from the outside to the inside of the symmetric bubble (dashed-black line 
in Figure \ref{fig:7}), since $\delta v(\xi) > 0$. 
However, once the symmetric bubble enters the region where $T(\xi) < T_c$, the velocity
$\delta v(\xi)$ becomes negative and the plasma inside the symmetric bubble, as seen from the symmetric bubble wall, 
moves towards the wall instead of away from it. Then, the plasma flows from the inside to the outside of the symmetric bubble 
(dashed-black line in Figure \ref{fig:7})
and the symmetric bubble starts contracting, shrinking progressively and eventually disappearing.

Given that a symmetric bubble expands to reach its maximal size for $\xi = \xi_c$ and then shrinks back to nothing, only plasma
volume elements whose trajectories $r_c(\tilde{r}_0,\xi)$ are close enough to the one of the plasma element in which the 
symmetric bubble nucleates, $r_c(r_0,\xi)$, will actually feel the presence of the symmetric bubble by going through it
(dashed-black line in Figure \ref{fig:7}). The maximum and minimum value of $\tilde{r}_0$ are, respectively

\be
\label{eq:growth4}
\tilde{r}_{0_{\mathrm{max}}} = r_0 + \delta r_0^{(2)} \quad \quad \quad  \tilde{r}_{0_{\mathrm{min}}} = r_0 - \delta r_0^{(1)}
\, ,
\ee

\noindent where $\delta r_0^{(2)}$ and $\delta r_0^{(1)}$ are obtained from the conditions (see Figure \ref{fig:7})

\be
\label{eq:growth5}
r_{\mathrm{out}}(r_0, \xi_c) = r_c(r_0 + \delta r_0^{(2)},\xi_c) \quad \quad \quad 
r_{\mathrm{in}}(r_0, \xi_c) = r_c(r_0 - \delta r_0^{(1)},\xi_c) 
\, .
\ee
Getting $\delta r_0^{(2)}$ and $\delta r_0^{(1)}$ as a function of $r_0$ and $\xi$ is rather 
involved; we present the calculation in Appendix B. 

So far we have focused only on  
the evolution of the symmetric bubble along the broken bubble's radial direction, 
disregarding the expansion of the symmetric bubble along the angular directions. The full three-dimensional problem, together with 
the departure from the initial spherical shape of the symmetric bubble as it evolves, are difficult matters to study 
(see Appendix B). However, given spherical symmetry it is a reasonable approximation to consider 
that the volume of plasma that goes through a symmetric bubble is roughly given by (\ref{B9})

\be
\label{eq:growth6}
V_S(r_0) \simeq (\delta r_0^{(1)} + \delta r_0^{(2)})^3 \equiv (\delta r_0)^3
\, .
\ee
This last result will be used in the next section, where we analyse the possibility of achieving 
enough baryogenesis through the nucleation of symmetric bubbles.

\section{Fast Bubble Expansion and EW Baryogenesis.}
\label{sec_6}
  
The standard EW baryogenesis mechanism is based on the interaction between the wall of the expanding bubbles 
and the plasma in front of it \cite{Cohen:1990py,Cohen:1991iu,Joyce:1994fu}, which leads to a 
CP-asymmetric reflection on the wall of certain particle species. These particle
asymmetries are subsequently diffused into the plasma in the symmetric phase just in front of the bubble wall \cite{Joyce:1994fu}, 
where sphalerons are active and capable of converting the CP asymmetry into a net baryon number. The generated baryon number is then  
carried into the broken phase (where it stays frozen) as the wall passes by. 

The diffusion process plays a key role in the generation of the baryon asymmetry, since it connects the non-equilibrium CP violating physics 
(occurring within the bubble wall) to the sphaleron baryon number violating processes (occurring 
in the symmetric phase outside the bubble). Therefore, in order for the whole mechanism to be able to 
generate the baryon asymmetry of the universe, the diffusion process has to be effective, meaning that the timescale for the 
diffusion of the asymmetries into the symmetric phase has to be smaller than the time the wall takes to sweep through the 
plasma just in front of it (suppressing the sphaleron processes as it passes through). This puts an upper 
bound on the relative velocity between the bubble wall and the plasma in the symmetric phase 
$v_r \lesssim D/L_{w} \sim 0.15 - 0.3$ (with $D$ a diffusion constant and $L_{w}$ 
the wall thickness) \cite{Joyce:1994fu}. Moreover, effective diffusion cannot take place if the relative velocity $v_r$ is greater 
than the speed of sound of the plasma $c_s$. This seems to prevent the generation of the baryon asymmetry of the universe via
EW baryogenesis (even if all three Sakharov conditions are present at the EW scale) for scenarios with
relatively fast bubble walls, and in particular when the bubbles expand as detonations, case in which 
the bubble walls are supersonic\footnote[11]{As explained in the introduction, while 
this is not the case for the MSSM, due to the phase transition being at most weakly 
first order and the large amount of friction on the wall \cite{John:2000zq}, it 
may indeed be the case for other extensions of the Standard Model with a stronger EW phase transition, 
such as the NMSSM \cite{Huber:2007vva} or the Standard Model extended with a 
singlet scalar \cite{EGKR,AndersonHall,Ham:2004cf,Profumo:2007wc,Espinosa:2011ax,Espinosa:1993bs,Espinosa:2007qk}.}.

Nevertheless, even if diffusion in front of the broken phase bubble wall 
is ineffective, and the EW baryogenesis mechanism previously described does not work, 
it may still be possible to generate the baryon asymmetry when the bubbles expand as detonations. 
Recalling the discussion of section \ref{sec_3}, the heating of the plasma behind the broken bubble wall caused by the detonation wave
may rise the temperature of a region behind the wall above $T_c$, and small bubbles of the symmetric phase will be nucleated
in this region (see section \ref{symmetric_nucleation2}), in whose interior sphalerons processes are unsupressed 
(even though the symmetric bubble
maximal size is always much smaller than the size of the broken bubble, as seen from Figure \ref{fig:5},
it is still larger than the characteristic size of a sphaleron configuration by many orders of magnitude). 
Since this small symmetric bubbles are nucleated at rest in the plasma, they move along with the 
plasma volume element in which they are nucleated, and therefore they go through the detonation wave. While they are in the region
where $T > T_c$, the symmetric bubbles will slowly expand, with the plasma flowing across their wall from the 
broken to the symmetric phase. However, the symmetric bubbles reach a maximum size and then start shrinking as they enter
the region where $T < T_c$, getting smaller until eventually disappearing (see section \ref{sec_5}). During the shrinking
process, the plasma flows across the symmetric bubble wall from the symmetric to the broken phase.  

\begin{figure}[ht]
\begin{center}
\includegraphics[width=0.62\textwidth, clip ]{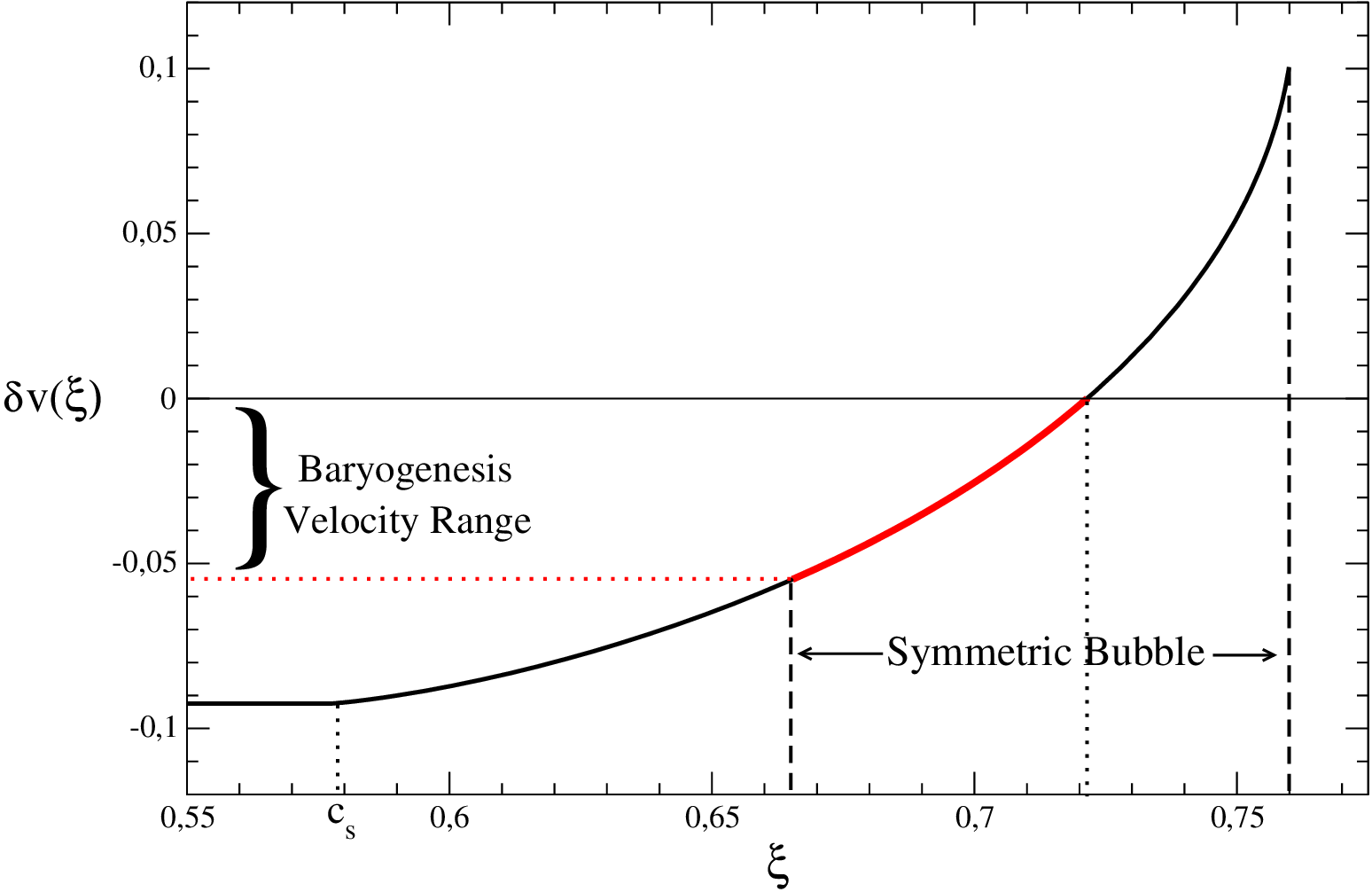}
\caption{\label{fig:8}
\small
$\delta v(\xi)$ for a nucleated symmetric bubble (for $v_w = 0.76$ and $\alpha_N = 0.078$). The red line
corresponds to $\xi < \xi_c$, so that $\delta v(\xi)$ becomes negative, the plasma flows from the symmetric
phase to the broken phase and baryogenesis can take place. We see that the range of velocities involved in this case
is $\left|\delta v\right| \sim 0.01- 0.05 \ll c_s$.}
\end{center}
\end{figure}

During the symmetric bubble expansion, EW baryogenesis is not possible, since the plasma flow across
the bubble wall goes from the broken to the symmetric phase and there is no 
way any baryon asymmetry generated by sphalerons can successfully
be transported into the broken phase before it is washed out. On the other hand, during the symmetric bubble contraction,
the plasma flows across the wall from the symmetric to the broken phase (just as for the expansion of 
the broken bubble) and any potential baryon asymmetry generated by sphalerons in the symmetric phase is automatically 
transported into the broken phase as the symmetric bubble wall passes through. Moreover, 
as opposed to the broken bubble (for which diffusion is suppressed due to $v_w > c_s$), for the symmetric bubbles
the relative velocity between the wall and the plasma in the symmetric phase as they contract 
is $\left|\delta v \right| \ll c_s$ (see (\ref{eq:growth1}) and Figure \ref{fig:8}), and so the diffusion 
of the particle asymmetries into the symmetric phase is indeed effective.

Let us now suppose that the theory beyond the Standard Model responsible for the 
first order EW phase transition has all the necessary ingredients 
for successfully generating the observed baryon asymmetry of the universe via EW baryogenesis (sufficient CP violation and 
phase transition sufficiently strong), but that the expansion  
of the broken phase bubbles is supersonic and 
proceeding via detonations. 
The EW baryogenesis mechanism can then still take place 
close to the surface of the symmetric bubbles. Since the symmetric bubble wall is expected to be in all respects similar to the 
broken bubble wall, the baryogenesis mechanism at the microscopic level 
is the same as in the standard case; therefore, it can operate due to the small expansion velocity of the symmetric bubbles. 
%

Still, the EW baryogenesis scenario presented here needs to address a potential ``filling factor" suppression: this is  
due to the fact that, depending on the number of nucleated symmetric bubbles, 
it is not guaranteed that all the plasma inside the broken bubble will have undergone
EW baryogenesis at one time or another by going through a symmetric bubble.
If we define $n_B$ to be the local baryon number density generated during the process of EW baryogenesis
and $\bar{n}_B$ to be the average baryon number density in the universe after the EW phase transition has completed,
then in the present baryogenesis scenario they are related by  

\be
\label{eq:EWBG1}
\bar{n}_{B} = 
n_{B} \, \frac{V^{\mathrm{Sym}}}{V^{\mathrm{B}}}
\, ,
\ee

\noindent where $V^{\mathrm{B}}$ is the volume of a broken bubble and $V^{\mathrm{Sym}}$ is the total volume of plasma within 
the broken bubble having undergone EW baryogenesis by going through the symmetric bubbles. We then define 
the filling factor $\Upsilon \equiv V^{\mathrm{Sym}}/V^{\mathrm{B}}$. Since
$V^{\mathrm{B}}$ and $V^{\mathrm{Sym}}$ ultimately depends on the phase transition parameters (such as $T_c$ or $T_N$)
and on the wall velocity $v_w$, (\ref{eq:EWBG1}) can be written as\footnote[12]{While not shown explicitly, 
$\Upsilon$ depends on other phase transition parameters apart from $T_c$.} 

\be
\label{eq:EWBG2}
\frac{\bar{n}_{B}}{ n_{B}} = \Upsilon (T_{c},v_w) \, .
\ee

In order to get a positive filling factor, a necessary condition is that $T_{-} > T_c$ (see Figure \ref{fig:3}). 
However, this is not a sufficient condition: one still needs to nucleate at least one symmetric bubble inside each 
broken bubble during the phase transition, assuming that the broken bubble 
was nucleated at the beginning of the phase transition (this assumption is well justified since those bubbles are 
in any case the ones that eventually cover most of the horizon volume at the end of the phase transition). This 
provides a lower bound on $T_{-}$ (as a function of $v_w$) for $\Upsilon > 0$. 
From (\ref{eq:nucleation5}), this condition reads

\be
\label{eq:EWBG3}
\mathcal{N}^{S}_{\textrm{Bubbles}}(\beta^{-1}) = 1 \, .
\ee

Moreover, since 
the volume of plasma going through one symmetric bubble is very small compared to the size of the 
broken bubble at any time, $V_S(t) \ll t^3$ (see (\ref{eq:growth6}) and the discussion in section \ref{sec_5}), it 
is only possible to obtain
a sizable value of the filling factor $\Upsilon \lesssim 1$ if a large number of 
symmetric bubbles are nucleated inside each broken bubble. If only a few 
symmetric bubbles are nucleated per broken bubble, then $\Upsilon \ll 1$ and $\bar{n}_B$ will be highly 
suppressed\footnote[13]{This is similar to the case of baryogenesis from cosmic strings, 
for which the baryon asymmetry produced is volume-suppressed \cite{Cline:1998rc} with respect to
$n_B$ due to the fact that the strings sweep just some small part of the volume of the universe as they move.}.  Note that given the present uncertainties in the 
computation of $n_B$ in EW baryogenesis scenarios \cite{Prokopec:2003pj,Cirigliano:2011di}, 
a reasonable requirement for successful EW baryogenesis is 
$\Upsilon > 0.1-0.5$. 

Let us now turn to the actual computation of the filling factor $\Upsilon = V^{\mathrm{Sym}}/V^{\mathrm{B}}$.  
Since the final radius of a typical broken bubble is $R_B \simeq v_w \, \beta^{-1}$, we have 

\be
\label{eq:EWBG4}
V^{\mathrm{B}} = \frac{4}{3} \, \pi \, v_w^3 \, \beta^{-3}
\, .
\ee
On the other hand, the total volume of plasma within a broken 
bubble having undergone EW baryogenesis $V^{\mathrm{Sym}}$ can be
written as 

\be
\label{eq:EWBG5}
V^{\mathrm{Sym}} = \int_{t_{\mathrm{min}}}^{\beta^{-1}} V_S(t)\, N^S_{\textrm{B}}(t) \, dt
\, ,
\ee

\noindent where $V_S(t) = \left(\delta r_0 (t)\right)^3$ is the volume of plasma that goes through a symmetric bubble,
obtained from (\ref{eq:growth6}) and (\ref{B8}), and $N^S_{\textrm{B}}(t)$ is the number of symmetric 
bubbles nucleated inside a broken one per unit time, given by (\ref{eq:nucleation3}). The lower integration 
limit $t_{\mathrm{min}}$, defined by the condition

\be
\label{eq:EWBG8}
\frac{\delta r_0 (t_{\mathrm{min}})}{v_w} \, N^S_{\textrm{B}}(t_{\mathrm{min}}) = 1
\, ,
\ee

\noindent represents the fact that one needs to nucleate at least one 
symmetric bubble during the time interval $\delta r_0 (t)/v_w$ for $V_S(t)$ to be nonzero
(and EW baryogenesis to happen during that time interval). 
Moreover, the more symmetric bubbles are nucleated
during a time interval $\delta r_0 (t)/v_w$, the less volume is left for further symmetric bubbles to nucleate,
and so the value of the integrand in (\ref{eq:EWBG5}) cannot be arbitrarily high, but is bounded by 

\be
\label{eq:EWBG6}
V_{S}(t) \, N^S_{\textrm{B}}(t)\, \frac{\delta r_0 (t)}{v_w} = 
N^S_{\textrm{B}}(t)\, \frac{\left(\delta r_0 (t)\right)^4}{v_w} = 
\frac{4}{3}\, \pi \, v_w^3 \left[t^3 - 
\left(t - \frac{\delta r_0 (t)}{v_w} \right)^3 \right] \, ,
\ee

\noindent where the r.h.s of (\ref{eq:EWBG6}) corresponds to the total volume a broken bubble sweeps in a time interval 
$\delta r_0 (t)/v_w$, and no more symmetric bubbles may be nucleated in that volume. 
Since $\delta r_0 (t) \propto t$ and $N^S_{\textrm{B}}(t) \propto t^3$, if the bound (\ref{eq:EWBG6}) 
is reached for $t = \tau < \beta^{-1}$, then (\ref{eq:EWBG5}) should
be replaced by 

\be
\label{eq:EWBG7}
V^{\mathrm{Sym}} \simeq \int_{t_{\mathrm{min}}}^{\tau} V_S(t)\, N^S_{\textrm{B}}(t) \, dt + \frac{4}{3}\, \pi \, v_w^3 
\left[\beta^{-3} - \tau^3\right]
\, .
\ee

It is now clear that, for a given model that produces a first order EW phase transition, it is possible to obtain
both $V^{\mathrm{Sym}}$ and $V^{\mathrm{B}}$ as a function of the parameters of the transition and the wall velocity $v_w$,
using the results from sections \ref{sec_3}, \ref{sec_4} and \ref{sec_5}. 
In the next section we will study the efficiency 
of the proposed EW baryogenesis mechanism (including the evaluation of the 
filling factor $\Upsilon$) in an explicit model.

\section{Supersonic Baryogenesis in a Specific Model.}
\label{sec_7}

The analysis of the heating of the plasma due to the detonation wave (section \ref{sec_3}) can be performed  
without relying on any particular symmetry breaking scenario, simply using the hydrodynamic analysis from section \ref{sec_2} and 
specifying an equation of state for the plasma. On the other hand, the analysis of the nucleation and expansion of symmetric bubbles
inside a broken bubble requires the knowledge of the free energy of the system $\mathcal{F}$, since the computation of the 
tunneling action $S_{3}$ can only be performed once the free energy has been specified. 
Consequently, in order to explore the viability of the EW baryogenesis mechanism here proposed, we concentrate on a specific scenario: the 
Standard Model extended by a set of singlet scalar fields $S_{i}$. As explained below, for the sake of simplicity we take a particular limit of this model where the analysis can be performed analytically, 
and discuss afterwards the possible implications of a more general situation.   

\subsection{Singlet Scalar Extension of the Standard Model.\label{sec_Potentialstrong}}

Extending the Standard Model by one or several singlet scalar fields coupled to the Higgs field yields a very attractive scenario in particular concerning EW baryogenesis, since this can easily give rise to a rather strong first order EW phase transition 
\cite{AndersonHall,Ham:2004cf,Profumo:2007wc,Espinosa:2011ax}. 
If one adds only one extra singlet field (either real or complex), the phase transition 
is sufficiently strong only if both the Higgs and the singlet
change their VEVs during the phase transition (see for example \cite{Profumo:2007wc,Espinosa:2011ax}).  
On the other hand, when the Higgs is the only field driving the phase transition (and for a Higgs mass
above the LEP bound $m_h > 114.4$ GeV \cite{LEP}) the presence of one extra singlet 
is not sufficient \cite{Espinosa:1993bs}, and a relatively large number of singlets $N_S$ is needed 
in order to achieve a strong phase transition \cite{Espinosa:2007qk}.

The tunneling action $S_{3}$ in a phase transition involving both the singlet field(s) and the Higgs field is 
 difficult to obtain, and in general cannot be found analytically (see \cite{Konstandin:2006nd}). Therefore, for simplicity we  
consider here the particular case in which the Higgs is the only field changing VEV during the phase 
transition. This allows us to use a semi-analytical solution for $S_{3}$. Then, in order to achieve a 
sufficiently strong first order phase transition, we have to add many singlet scalar 
fields (we take $N_S = 8$). We start from the following tree-level scalar potential 

\be
\label{eq:Vtree}
V_{\rm tree} = m^2 H^\dagger H + \lambda (H^\dagger H )^2
+ \zeta \sum_i  H^\dagger H \, S_i^2\ ,
\ee

\noindent where for simplicity we assume that the mass term $m_{S_i}^2 S_i^2$ is absent, and we take an universal coupling between the Higgs and the singlet scalars. 

Since in the above model (\ref{eq:Vtree}) 
the scalar singlet fields do not take a VEV, we can concentrate on the potential along the Higgs field
direction. In the background Higgs field configuration defined by $\left\langle H^0 \right\rangle = h/\sqrt{2}$, 
the effective potential at 1-loop in Landau gauge and $\overline{MS}$ renormalization scheme is 

\be
\label{eq:V1-loop}
V_{\rm 1-loop} = \frac{m^2}2 h^2 +
\frac\lambda4 h^4 + \sum_\alpha \frac{N_\alpha M_\alpha^4(h)}{64 \pi^2}
\left[ \mathrm{log} \left(\frac{ M_\alpha^2(h)}{Q^2}\right)- C_\alpha \right]\ .
\ee

The subscript $\alpha =\{Z, W, t, S_i \}$ denotes the gauge
bosons ($Z^0$ and $W^\pm$), top quark and singlet scalar fields with $N_\alpha =
\{3, 6, -12, N_S \}$ (for the quantitative analysis, we take $N_S = 8$), 
while $C_\alpha = 5/6$ ($3/2$) for gauge bosons (fermions and scalars). We are neglecting
the contributions from all the quarks and leptons except the top
quark, and also from the Higgs itself and the Goldstone bosons. The $h$-dependent tree-level masses
are
\be
M^2_{S_i}(h) = \zeta h^2\ , \quad 
M^2_Z(h)=\frac{1}{4}( g^2 + 
g^{\prime2}) h^2\ , 
\quad M^2_W(h)= \frac{1}{4} g^2 h^2 \ , \quad M^2_t(h)= \frac{1}{2} y_t^2 h^2\ ,
\ee
where $g$ and $g^\prime$ are the SM gauge couplings and $y_t$ the
top quark Yukawa coupling. The renormalization scale $Q$ is chosen to be the top mass $Q = M_t(v)$,
and the parameters at that scale are fixed to recover $\left\langle h \right\rangle = v = 246$ GeV. 

At a finite temperature $T$, the free energy of the Higgs field at 1-loop is given by

\bea
\label{eq:FreeEnergy1}
\mathcal{F}(h, T) =  V_{\rm 1-loop}(h) + \sum_\alpha \frac{N_\alpha T^4}{2 \pi^2} 
\int_0^\infty dk \, k^2 \,\mathrm{log} \left(1 \pm e^{-\sqrt{k^2 + M^2_{\alpha}(h)/T^2}}\right)
\ .
\eea
If we consider the high temperature limit $T \gg M_{\alpha}(h)$, we can perform an expansion in $M^2_{\alpha}(h)/T^2$ in 
(\ref{eq:FreeEnergy1}), obtaining  

\bea
\label{eq:FreeEnergy2}
\mathcal{F}(h, T) =-\frac{\pi^2\, g^{*}}{90} \, T^4 + \frac{m^2}{2} h^2 +
 \frac{\lambda}{4} h^4 + \sum_\alpha \frac{N_\alpha M_\alpha^4(h)}{64 \pi^2}
\left[ \mathrm{log} \left(\frac{a_\alpha \, T^2}{Q^2} \right)- C_\alpha \right] + \nonumber \\ 
\sum_{\alpha \in \textrm{bosons}} N_\alpha 
\left(\frac{T^2 \, M^2_\alpha(h)}{24} - \frac{T \, M^3_\alpha(h)}{12\pi} \right) - \sum_{\alpha \in \textrm{fermions}}  
 \frac{N_\alpha \, T^2}{48}\, M^2_\alpha(h)\ ,
\eea

\noindent where $g^{*} = \sum \left(g_{b} + (7/8) g_{f}\right)$ now includes also the contribution 
to the number of relativistic degrees of freedom of all the quark and leptons, as well as the 
Higgs and Goldstone bosons. Also, $\mathrm{log} (a_\alpha) = 5.41$ ($2.64$) for bosons (fermions). 

Note that both in (\ref{eq:FreeEnergy1})
and (\ref{eq:FreeEnergy2}) we are not including the contribution
to the free energy of the Higgs field coming from the resummation of bosonic zero modes (the so-called ``Ring" or ``Daisy" 
contribution to $\mathcal{F}$). This contribution is potentially important, specially at high temperature, and in a complete study of the phase transition it should be included; here we neglect it since we want to be able to perform an analytic study. 
Due precisely to the fact that we are ignoring this contribution, and that we are 
neglecting the contributions from the Higgs and Goldstone bosons in (\ref{eq:V1-loop})
and (\ref{eq:FreeEnergy1}), we can write (\ref{eq:FreeEnergy2}) as

\be
\label{eq:FreeEnergy3}
\mathcal{F}(h, T) = A(T)\, h^2 - B(T) \,h^3 + \lambda_T(T)\, h^4 + \, \mathrm{(field \,\, independent \, \, terms)}\ ,
\ee

\noindent with

\bea
\label{eq:FreeEnergy4}
A(T) = \frac{m^2}{2} + \sum_{\alpha \in \textrm{bosons}} 
\frac{N_\alpha \,  T^2 \, M^2_\alpha(v)}{24\, v^2}  - \sum_{\alpha \in \textrm{fermions}}  
 \frac{N_\alpha \, T^2\, M^2_\alpha(v)}{48\, v^2} \nonumber \\
B(T) = \sum_{\alpha \in \textrm{bosons}} \frac{N_\alpha\, T \, M^3_\alpha(v)}{12\, \pi\, v^3} \quad \quad \quad 
\quad \quad \quad \quad \quad \quad \quad \quad \nonumber \\
\lambda_T(T) = \frac{\lambda}{4} +  \sum_\alpha \frac{N_\alpha M_\alpha^4(v)}{64 \,\pi^2\, v^4}
\left[ \mathrm{log} \left(\frac{a_\alpha \, T^2}{Q^2} \right)- C_\alpha \right]  \, \, \, \quad \quad \quad \nonumber
\eea
Arranging (\ref{eq:FreeEnergy2}) in the form (\ref{eq:FreeEnergy3}) makes the analysis much easier, as we now show. Eq.~(\ref{eq:FreeEnergy3}) describes a system with a first order phase transition, the $h^3$-term being 
responsible for the potential barrier between minima. The extrema of (\ref{eq:FreeEnergy3}) are: 

\be
\label{eq:Weak3}
h = 0 \quad \quad \quad \quad h_{\pm} = \frac{3 \,B(T)}{8 \, \lambda_T(T)} \pm 
\sqrt{\left(\frac{3\, B(T)}{8\, \lambda_T(T)}\right)^2-\frac{A(T)}{2 \,\lambda_T(T)}} \, .
\ee
The behaviour of the free energy (\ref{eq:FreeEnergy3}) with temperature is the following: at very high temperatures 
the free energy has a parabolic-like shape, with $h = 0$ being its only real extremum (a minimum), while
$h_{\pm}$ are complex. As the temperature lowers, it eventually reaches $T = T_{*}$, at which 
the term inside the square-root in (\ref{eq:Weak3}) vanishes and the extrema $h_{\pm}$ become real

\be
\label{eq:Weak4}
h_{+}(T_{*}) = h_{-}(T_{*}) = \frac{3 \,B(T_{*})}{8 \, \lambda_T(T_{*})} \quad \quad 
\quad \quad \frac{A(T_{*}) \, \lambda_T(T_{*})}{B^2(T_{*})} = \frac{9}{32} \, .
\ee
Below $T_{*} $, the free energy has two real minima ($h = 0$ and $h_{+} \neq 0$, with $h = 0$ still being the absolute minimum) 
separated by a potential barrier (its maximum being at $h_{-}$). As the temperature 
keeps lowering, it reaches $T_c$, the temperature at which the two minima have equal free energy 
$\mathcal{F}(h = 0, T_c)  = \mathcal{F}(h_{+}(T_c),T_c)$

\be
\label{eq:Weak5}
\frac{A(T_{c}) \, \lambda_T(T_{c})}{B^2(T_{c})} = \frac{8}{32} = \frac{1}{4}\, .
\ee
For $T < T_{c}$ the absolute minimum is $h_{+}$, and it becomes possible
to tunnel from $h = 0$ to $h_{+}$. In general, the computation of the tunneling action
$S_{3}$ has to be done numerically, but for the simple case of quartic potentials of a single scalar field 

\be
\label{eq:Vquartic}
V(\phi) = \lambda \phi^4 - b\, \phi^3 + a\, \phi^2  \quad \quad \quad \lambda, a ,b > 0
\ee

\noindent there exists a semi-analytic solution \cite{S3semianalytic} for the three-dimensional Euclidean action $S_{3}$, given by

\be
\label{eq:semianalitic}
S_{3} = \frac{\pi \,b}{\lambda^{3/2}} \frac{8 \,\sqrt{\delta}}{81\,(2-\delta)^2} \left(\beta_1 
\delta + \beta_2 \delta^2 + \beta_3 \delta^3\right) \quad \quad \quad \delta \equiv \frac{8 \, \lambda \,a}{b^2}
\ee

\noindent where $\beta_1 = 8.2938 $, $\beta_2 = - 5.5330 $, $\beta_3 = 0.8180$. 
Since Eq.~(\ref{eq:FreeEnergy2}), rewritten as in (\ref{eq:FreeEnergy3}), 
is precisely of the form (\ref{eq:Vquartic}) (dropping field independent terms), 
we can use Eq.~(\ref{eq:semianalitic}) to compute the tunneling action. Then, the temperature $T_N$ at which 
bubbles of the broken phase begin to nucleate is easily obtained from (\ref{eq:TN1}) 
and (\ref{eq:TN2}).

\begin{figure}[ht]
\begin{center}
\includegraphics[width=0.47\textwidth]{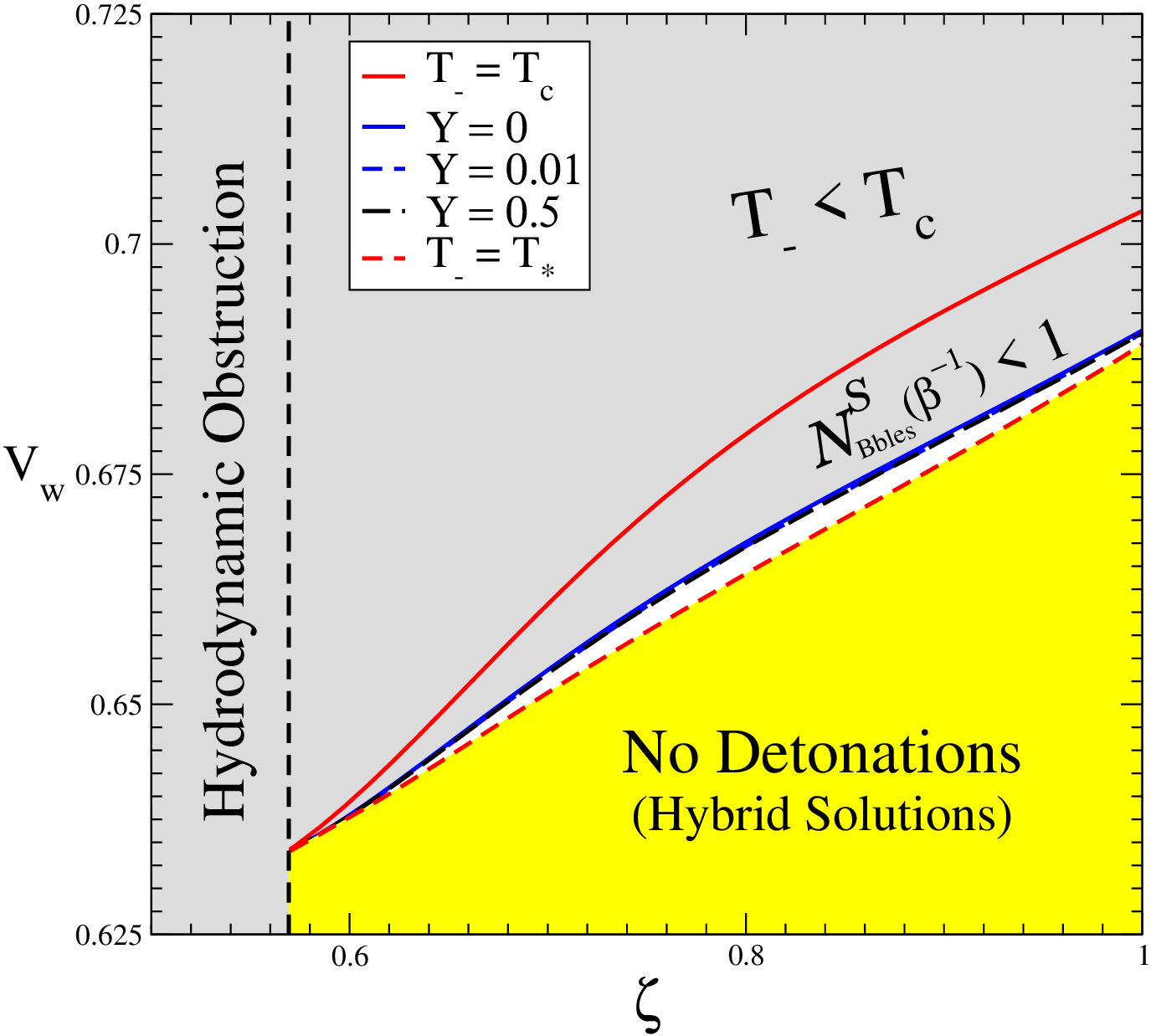} \hspace{2mm}
\includegraphics[width=0.47\textwidth]{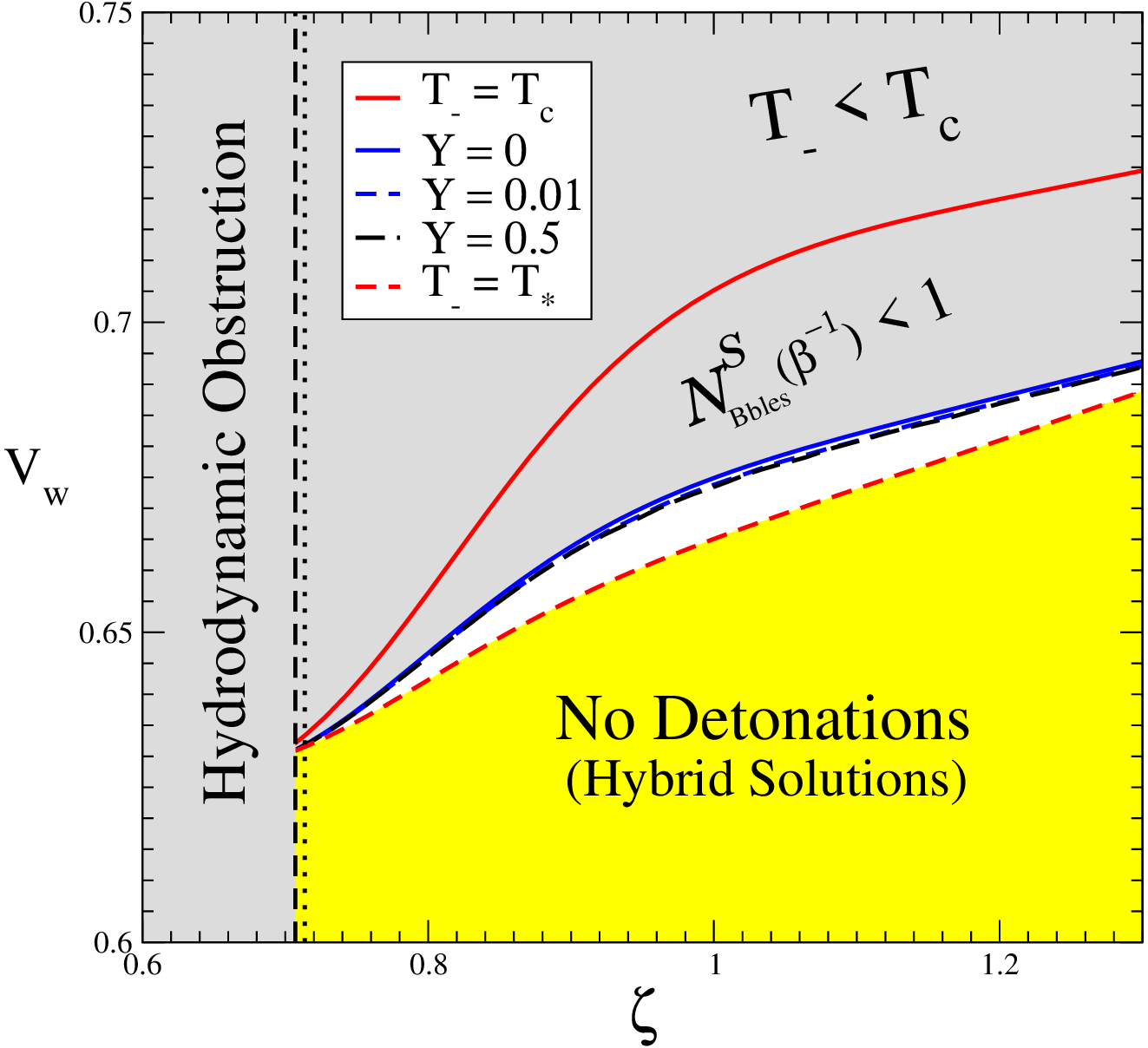}
\vskip 0.5 cm
\includegraphics[width=0.47\textwidth]{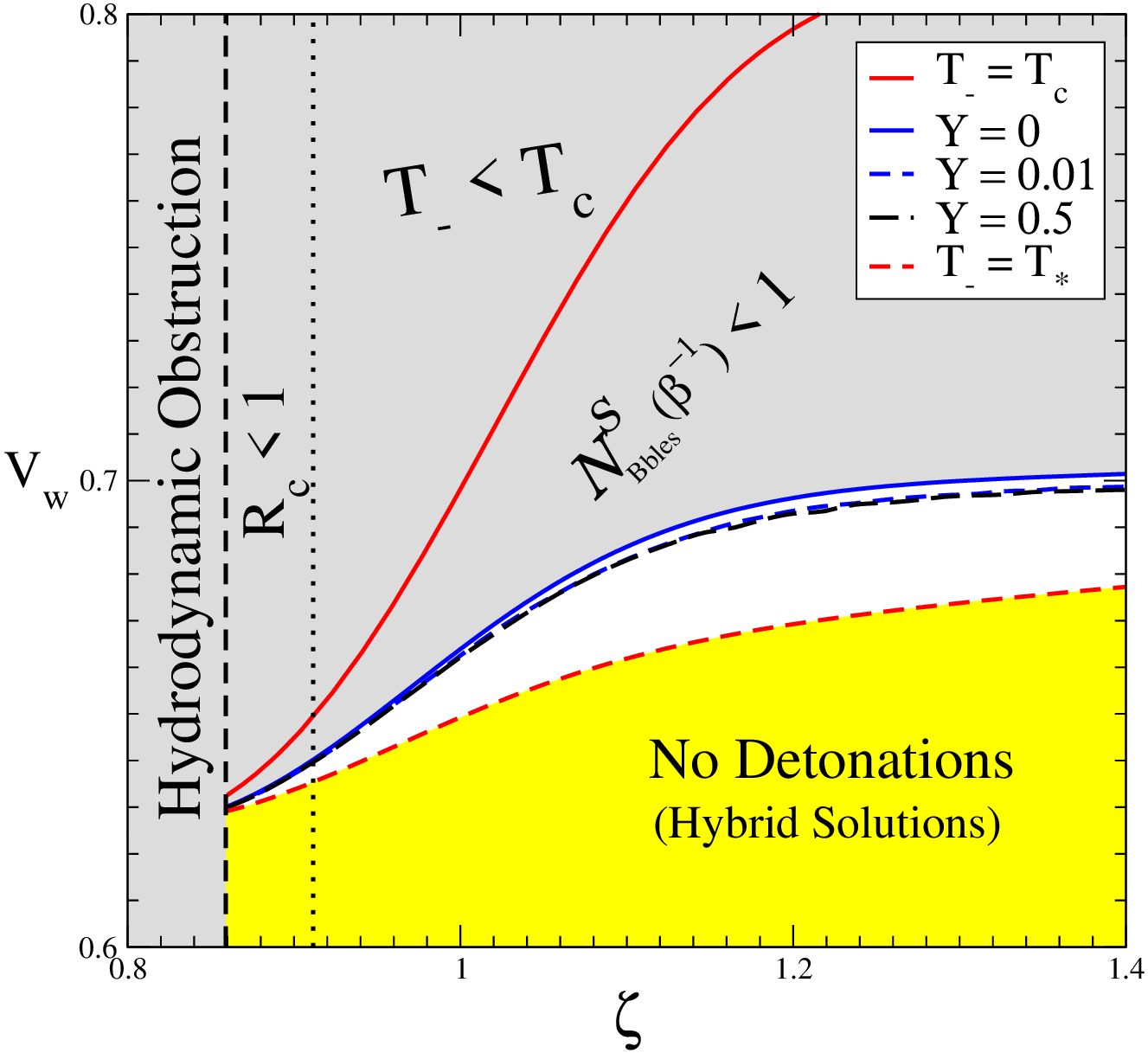} \hspace{2mm}
\includegraphics[width=0.47\textwidth]{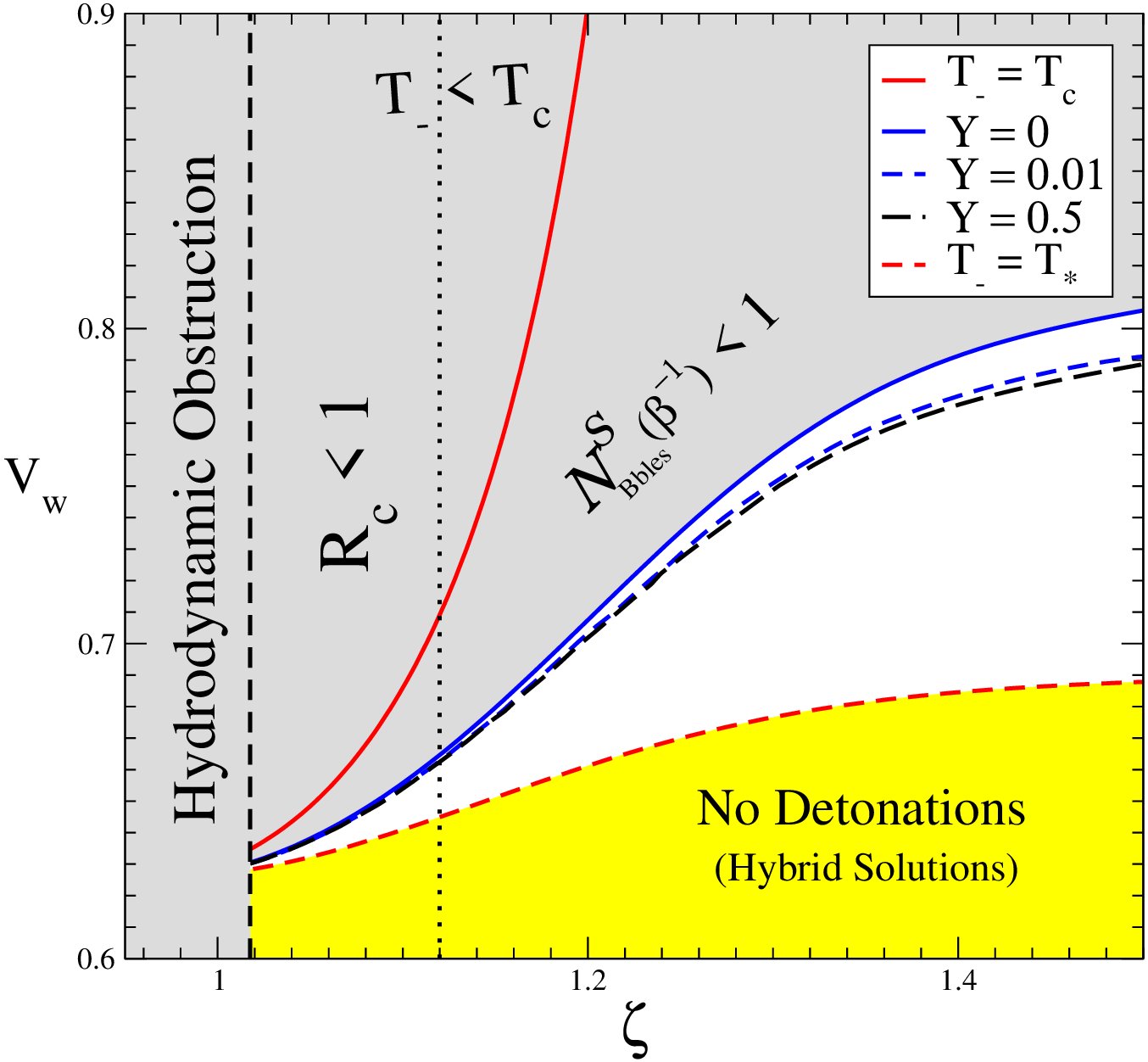}
\caption{\label{fig:11}
\small \small 
Allowed region for supersonic EW baryogenesis (in white) for the free energy (\ref{eq:FreeEnergy2}) 
with $m_h = 125$ GeV (up-left plot), $m_h = 150$ GeV (up-right plot), $m_h = 175$ GeV (down-left plot) 
and $m_h = 200$ GeV (down-right plot). In the yellow region, the stationary stable solutions are not detonations, but
hybrids (see main text).}
\end{center}
\end{figure}

If the nucleated bubbles expand as detonations, the condition (\ref{eq:Heating}) determines the region of parameter space 
where the temperature just behind the bubble wall $T_{-}$ is larger than $T_c$. Note that now $a_-$ and $\alpha_N$ are obtained 
from $\mathcal{F}$ \cite{EKNS}.
For the particular case we are analysing, once we fix the Higgs mass
$m_h$, the strength of the first order phase transition is fully controlled by the value of the 
coupling $\zeta$ between the Higgs and the scalar singlets (see Eq.~(\ref{eq:Vtree})). The larger is $\zeta$, the stronger is the phase transition. We present our results letting $\zeta$ and $v_w$ as free parameters. In Figure \ref{fig:11} we show the region in the ($\zeta, v_w$)-plane where
$T_{-}$ overcomes $T_c$, for several values of $m_h$ and bearing in mind
that a phase transition sufficiently strong for viable EW baryogenesis needs $R_c \equiv h_{+}(T_c)/T_c > 1$.
As previously stressed, however, $T_{-} > T_c$ is a necessary condition for the supersonic EW
baryogenesis mechanism to operate but not a sufficient one.

For $T_c < T_{-} < T_{*}$ we can compute the tunneling action from the broken to the symmetric minimum $S^S_{3}(T)$. 
This can be done by performing the shift $h = h_+ - (h_+ - h) = h_+ -\delta h$ in Eq.~(\ref{eq:FreeEnergy3}),
which becomes then $\mathcal{F}(h, T) = \mathcal{F}(h_+, T) + \mathcal{F}(\delta h, T)$, with

\bea
\mathcal{F}(\delta h, T) & = &\lambda_{T}(T)\, (\delta h)^4 - \left[ 4 \,\lambda_{T}(T) \, h_+ - B(T)\right] (\delta h)^3 +  \nonumber \\
 \nonumber \\
 & + & \left[A(T) - 3\, B(T) \, h_+ + 6 \, \lambda_{T}(T) \,h^2_+ \right] (\delta h)^2 \,.
\eea
Again, this potential is of the form (\ref{eq:Vquartic}), allowing us to use (\ref{eq:semianalitic}) to 
compute $S^S_3 (T)$. 
Once we know the Euclidean three-action, we can use the results of sections \ref{sec_4}, \ref{sec_5}
and \ref{sec_6} to obtain the region in the ($\zeta, v_w$)-plane where symmetric bubbles are nucleated
($\mathcal{N}^{S}_{\textrm{Bubbles}}(\beta^{-1}) \geq 1$) and thus the filling factor is non-zero $\Upsilon > 0$  
(this region lies obviously inside the one for which $T_{-} > T_c$). We then compute the filling factor
$\Upsilon$ in this region. The results are depicted in Figure \ref{fig:11}, 
which shows, for several values of the Higgs mass $m_h$, the region in the ($\zeta, v_w$)-plane where
the proposed EW baryogenesis mechanism operates (in white).

For values of $v_w$ below the dashed-red line in Figure \ref{fig:11} (yellow region), no stable 
detonation solutions exist, since above $T_*$ the broken minimum ceases to exist. In this region, 
the solution most likely develops a compression wave in front of the wall and turns into a hybrid\footnote[14]{
Hybrid solutions are usually allowed together with detonations \cite{EKNS}, even though typically
the pure detonation solution is selected by the system as its stationary expansion mode \cite{Ignatius:1993qn,KurkiSuonio:1995pp}. 
However, since in the region below the dashed-red line in Figure \ref{fig:11} 
a detonation is not stable, the hybrid solution may be realized in this case.} (or supersonic
deflagration \cite{KurkiSuonio:1995pp}), for which the heating of the plasma in the broken phase is typically lower \cite{EKNS}. 
It is possible to perform the analysis of the viability of the supersonic EW baryogenesis mechanism 
in the case of hybrid solutions, but this is left for future work. 

\begin{figure}[ht]
\begin{center}
\includegraphics[width=0.465\textwidth]{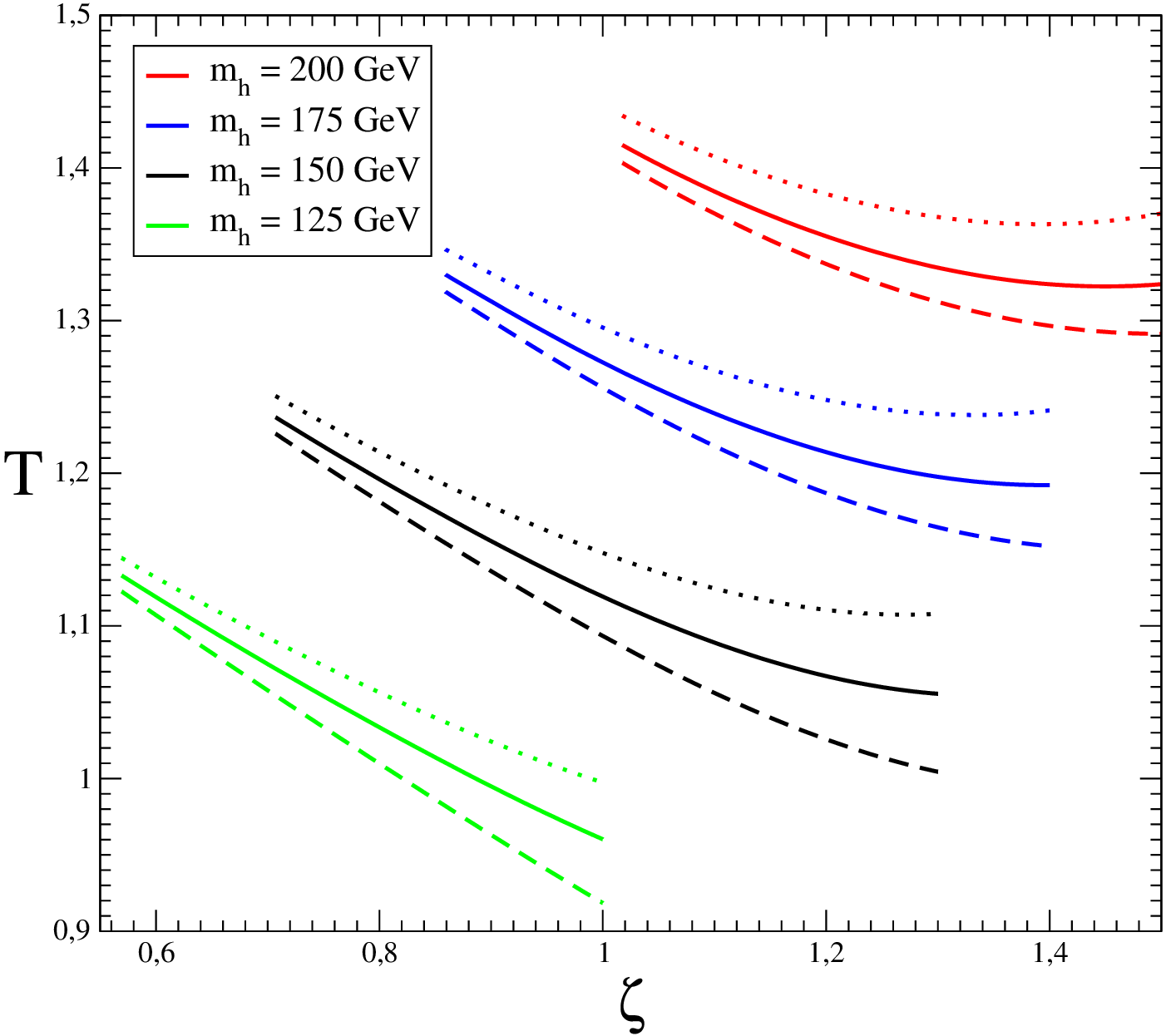} \hspace{2mm}
\includegraphics[width=0.485\textwidth]{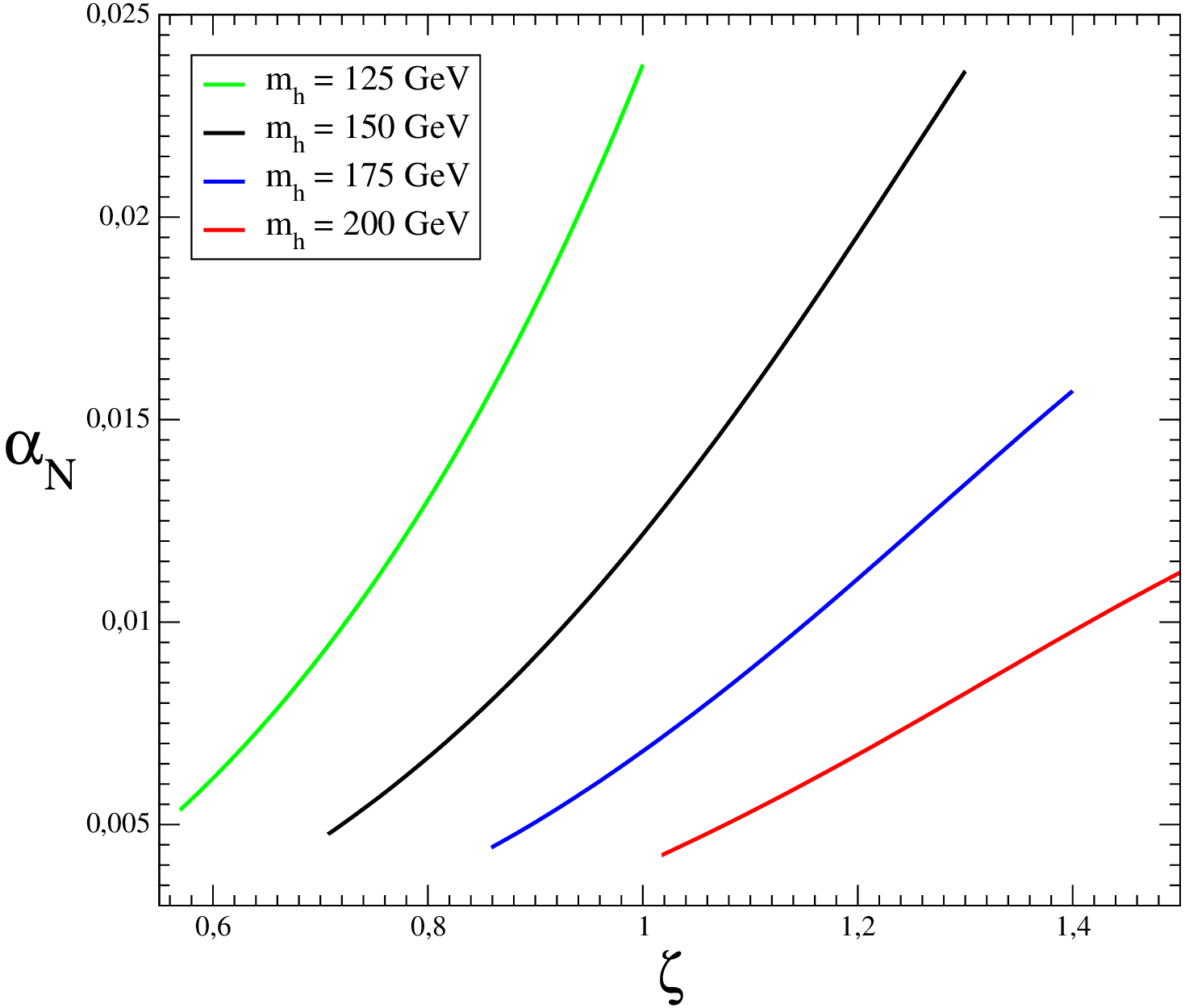}
\vskip 0.5 cm
\includegraphics[width=0.465\textwidth]{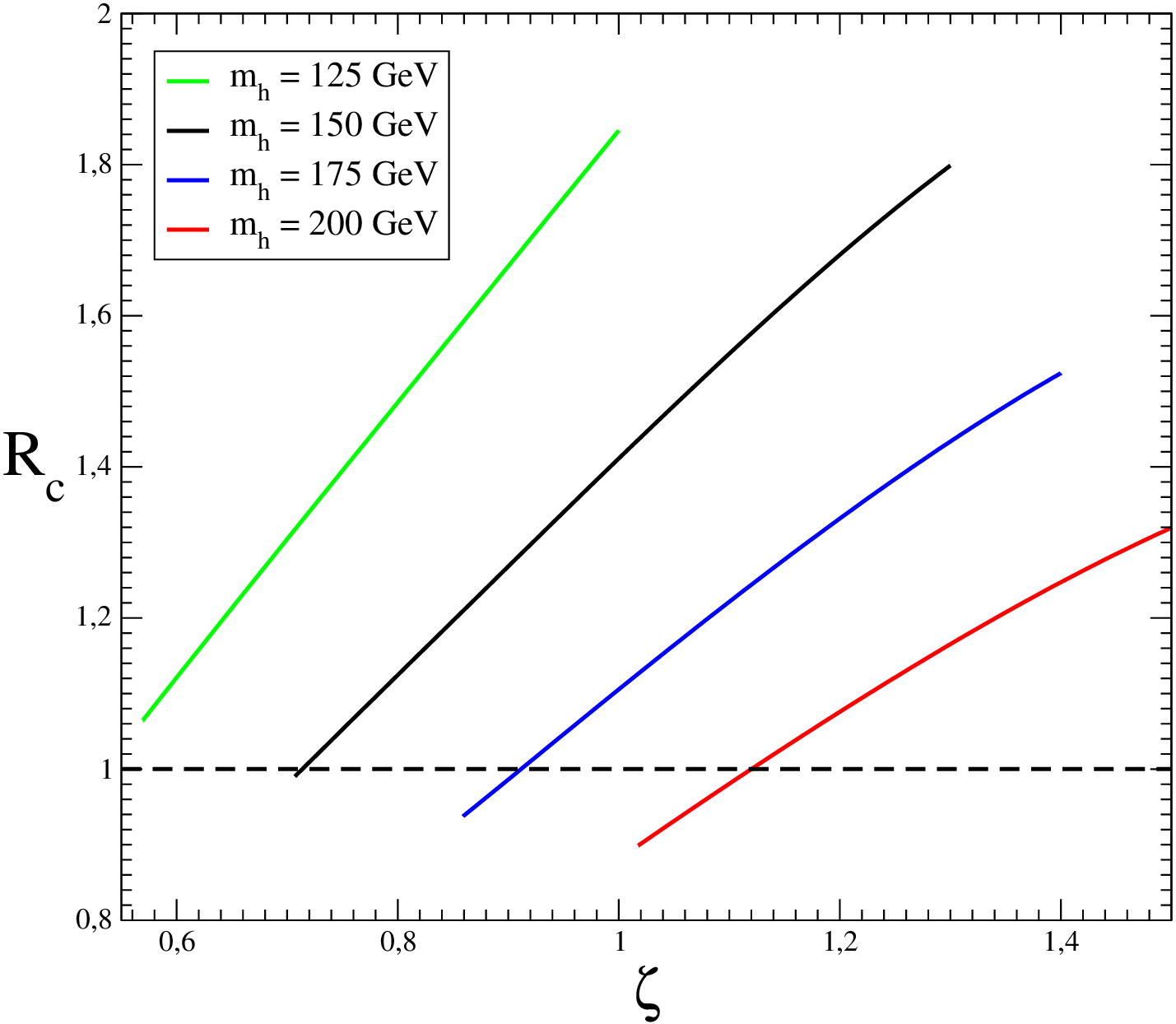} \hspace{2mm}
\includegraphics[width=0.49\textwidth]{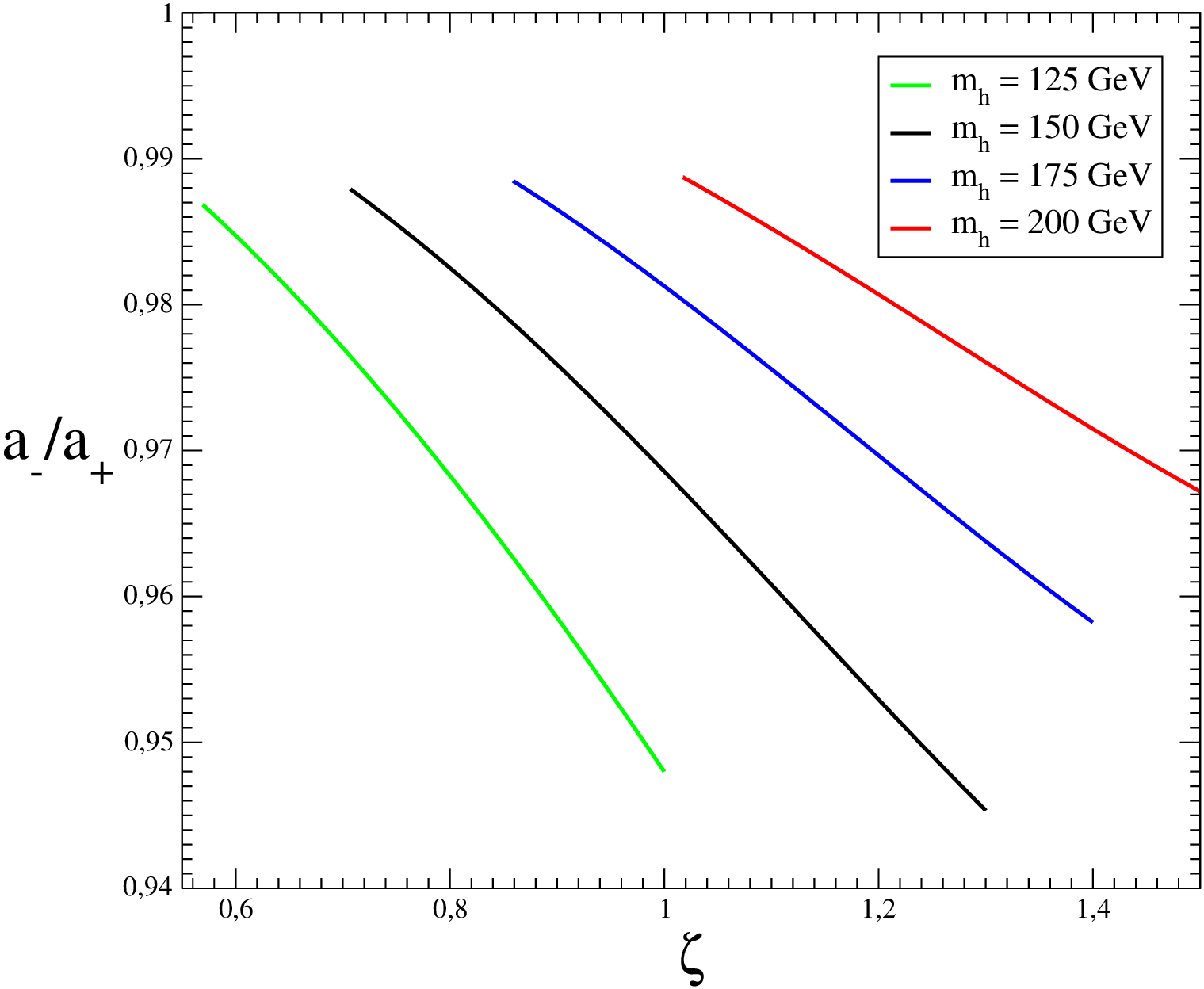}
\caption{\label{fig:12}
\small \small 
Dependence of various parameters of the phase transition with the coupling $\zeta$ for different 
values of $m_h$. Up-left: $T_N$ (dashed line), $T_c$ (solid line) and $T_*$ (dotted line). 
Up-right: $\alpha_N$. Down-left: $R_c \equiv h_{+}(T_c)/T_c$. Down-right: $a_{-}/a_{+}$. For 
each value of $m_h$, the range of $\zeta$ for which $T$, $\alpha_N$, $R_c$ and $a_{-}/a_{+}$ are 
plotted corresponds to the one chosen for the $\zeta$-axis in Figure \ref{fig:11}.}
\end{center}
\end{figure}

As seen from Figure \ref{fig:11}, in the present scenario both the $v_w$-ranges 
for which $T_{-} > T_c$ and for which $\Upsilon > 0$ become larger for increasing Higgs mass $m_h$. Since increasing Higgs mass weakens the phase transition, this might seem counterintuitive. However, for a fixed $m_h$, the $v_w$-ranges become larger 
as $\zeta$ increases, also meaning that the phase transition gets stronger. We conclude that the values of $m_h$ and $\zeta$,
being independent parameters, have a strong influence on both the strength of the phase 
transition (as measured by $R_c \equiv h_{+}(T_c)/T_c$) and on the $v_w$-range for which $\Upsilon > 0$; the interplay among these two parameter is complicated and it is possible to find values of $m_h$ and $\zeta$ with the same $R_c$ 
but very different ($\Upsilon > 0$, $v_w$)-range, as shown in Figure \ref{fig:12} (down-left plot).

This effect is similar to the one encountered in the analysis of section \ref{sec_3} for the case of the
Bag EoS, where we saw that the ($T_{-} > T_c$)-region becomes larger as $a_{-}/a_{+}$ decrases (for stronger phase transitions), 
but this region also becomes smaller as $\alpha_N$ increases (meaning also for stronger phase transitions). 
In the present case, $\alpha_N$ and $a_{-}$ both depend
on $\zeta$ and are not independent (see Figure \ref{fig:12}), resulting in the ($T_{-} > T_c$)-region
becoming larger as $\zeta$ increases and the phase transition gets stronger. However, $m_h$ and $\zeta$
are indeed independent, and the relation between the strength of the phase transition and
the size of the ($\Upsilon > 0$, $v_w$)-range happens to be opposite for them. In Figure \ref{fig:12} we 
show the dependence of various parameters of the phase transition
($T_c$, $T_N$, $T_*$, $\alpha_N$, $R_c$ and $a_{-}/a_{+}$) with the coupling $\zeta$ for different choices of 
$m_h$.  

\begin{figure}[ht]
\begin{center}
\includegraphics[width=0.55\textwidth]{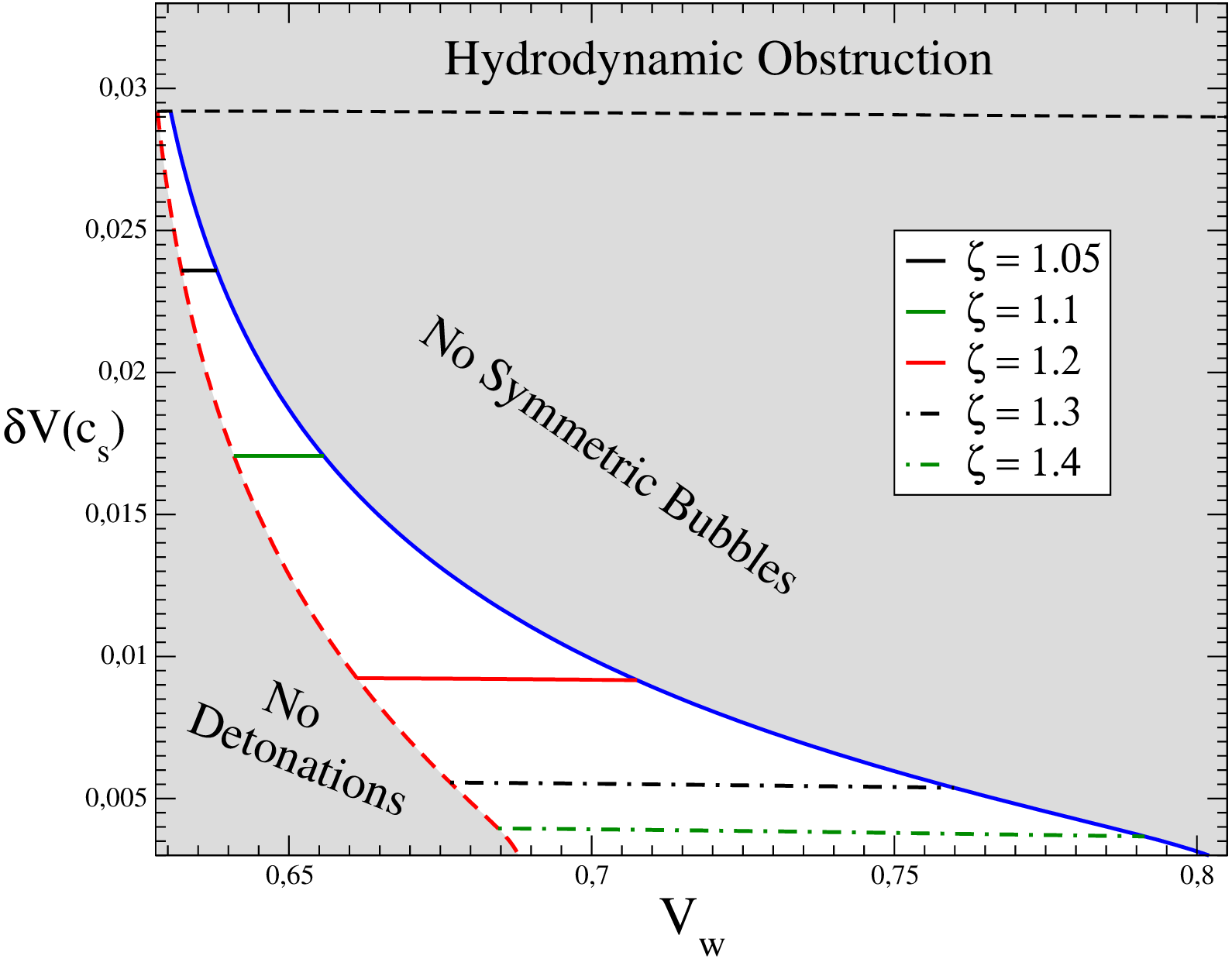}
\caption{\label{fig:13}
\small \small 
The relative velocity between the symmetric bubble wall and the surrounding 
plasma $\delta v$ as a function of the bubble wall velocity $v_w$ for different 
values of $\zeta$, and $m_h = 200$ GeV. The solid-blue and dashed-red lines correspond to the 
same lines in Figure \ref{fig:11}, namely $\Upsilon=0$ and $T_-=T_*$ (they therefore delimitate the region in which baryogenesis can occur).}
\end{center}
\end{figure}

Figure \ref{fig:13} shows the absolute value of the maximum\footnote[15]{The average relative velocity is 
typically a factor $2-3$ smaller, see Figure \ref{fig:8}.} possible relative velocity between the plasma 
and the symmetric bubble wall, corresponding to $\xi = c_s$, in the ($\zeta, v_w$)-plane for the particular 
case of $m_h = 200$ GeV. We see that typically $\left|\delta v (c_s) \right| \sim 10^{-3}-10^{-2} \ll c_s$, and 
diffusion is indeed efficient in the present scenario. As already discussed in section \ref{sec_6}, we expect this to be 
a general feature for any model leading to supersonic bubble expansion and inverse tunneling. 

Before concluding this section, let us briefly comment on the fact that part of the range for $m_h$ 
analyzed here ($m_h \in \left[ 140\right.$ GeV - $200$ GeV$\left. \right]$) has been recently 
excluded by LHC for the case of a Standard Model Higgs \cite{Aad:2011qi,ATLASCMS}. In singlet scalar field extensions of the
Standard Model it is possible to evade this constraint if the singlet scalars are light enough 
to open the decay channel $h \rightarrow S_i\,S_i$. 
In the present case, due to the fact that 
the couplings of the scalar singlets were chosen to be universal $\zeta_i \equiv \zeta$, the singlets are not light enough, since $m^2_S = \zeta v^2$ and $\zeta$ has to be
relatively large for the phase transition to be sufficiently strong. However, the coupling universality 
$\zeta_i \equiv \zeta$ is just a simplifying assumption, and it may be dropped (keeping $\sum_i N_i \zeta_i = N_S \zeta$) 
without changing the results of this section. Then, the constraint is easily evaded if at least one of the couplings $\zeta_i$
satisfies 

\be
\zeta_i < \frac{m^2_h}{4\, v^2}\,.
\ee

Finally, in the general case where both the Higgs field and the singlet(s)
change their VEVs during the transition (and after including all the 
relevant contributions to the free energy $\mathcal{F}$, 
such as the resummation of daisy diagrams), it is found that the phase transition is generically
stronger \cite{Profumo:2007wc,Espinosa:2011ax} than in the present scenario. 
We therefore expect that the supersonic EW baryogenesis mechanism
may also work in the more general case, although a more detailed study is needed.  

\section{Conclusions.}
\label{sec:Conclusions}

The expansion velocity of the broken phase bubbles during a first order EW 
phase transition plays a fundamental role both for EW baryogenesis and for the 
production of a stochastic background of gravitational waves. Whereas fast bubble walls 
are needed to produce a gravitational wave background detectable by future space-based 
interferometers, the standard EW baryogenesis mechanism is efficient only 
for scenarios in which the wall velocity is subsonic and small, since otherwise the diffusion
of particle asymmetries in front of the wall is strongly suppressed. So far, an explicit 
computation of the wall velocity during a first order EW phase transition has been performed only
in the case of the Standard Model (leading to $v_w \sim 0.35 - 0.45$ and for very small Higgs masses, 
already excluded by LEP), and in the case of a few specific models corresponding to weakly first order phase 
transitions (leading as well to rather low values for the wall velocity, $v_w \sim 0.05 - 0.1$ for the MSSM). Nevertheless, for  
scenarios where the EW phase transition would be more strongly first order, qualitative arguments 
already show that the wall velocity may be much larger, thus potentially rendering the standard EW
baryogenesis mechanism inefficient in those scenarios, even if the three Sakharov conditions for baryogenesis are satisfied.

Here we have presented a new mechanism for baryogenesis which operates for supersonic bubble walls, making it possible
to generate a sizable baryon asymmetry when the wall velocity of the expanding bubbles is large.
The mechanism is based on the fact that, when broken phase bubbles expand as supersonic detonations,
the plasma in the broken phase gets heated up in a small region behind the bubble wall. 
If the temperature in this region exceeds the critical temperature $T_c$, 
bubbles of the symmetric phase can nucleate and grow inside the broken phase bubble, sphalerons being reactivated inside 
these symmetric bubbles due to the restoration of EW symmetry. Since the symmetric bubbles are nucleated at rest in the 
moving broken phase plasma, they move through the detonation wave as they expand, entering at some point 
the region where the heating is not very efficient and the temperature is lower than $T_c$. 
In this region, the expansion of the symmetric bubbles is not any 
longer energetically favourable, and the bubbles start to shrink, the plasma then flowing from the symmetric to the broken 
phase across the symmetric bubble wall. Baryogenesis then occurs inside the symmetric bubbles via 
the usual diffusion mechanism (from standard EW baryogenesis),   
and the generated baryon asymmetry is naturally transferred to the broken phase as the 
symmetric bubble shrinks. We have shown that this 
process happens at a sufficiently low velocity to guarantee that diffusion is indeed effective. 


For this supersonic EW baryogenesis mechanism to work, one further requirement is that a large 
enough volume of the plasma goes through the symmetric phase. In order to see whether this can indeed be the case, 
we have chosen a specific model 
where the EW phase transition is strongly first order (the Standard Model coupled to several singlet scalar fields via the 
Higgs portal). We have obtained the region in parameter space (in terms of the universal coupling $\zeta$ between the Higgs and the 
singlet fields, the wall velocity $v_w$ and the Higgs mass $m_h$) where the filling factor $\Upsilon $ is large enough, and EW 
baryogenesis can occur for supersonic broken phase bubble walls. We have found that the region
in which the filling factor is large ($\Upsilon \rightarrow 1$) becomes smaller as $m_h$ decreases, 
while for a fixed $m_h$ it becomes larger for stronger phase transitions.

The process presented here can therefore lead to significant baryogenesis when the wall velocity $v_w$ is supersonic, which
may be the case in models where the EW phase transition is strongly first order. It could also accommodate a sizable GW 
signal with effective EW baryogenesis, leading in principle to a scenario
which can explain the baryon asymmetry of the universe and at the same time be tested with GW detectors.

\section*{Acknowledgments}

We thank Jose Ramon Espinosa, Thomas Konstandin and Geraldine Servant for their help and valuable comments
on the manuscript. We also thank Ruth Durrer, Mikko Laine and Tomislav Prokopec for interesting discussions. 
J. M. N. is supported by the European Commission under contract
PITN-GA-2009-237920 and by the Agence Nationale de la Recherche.

\section*{A. $\ $Matching Equations for Symmetric Bubble Walls}
\addcontentsline{toc}{section}{A $\ $Matching Equations for Symmetric Bubble Walls.}
\setcounter{equation}{0}
\renewcommand{\theequation}{A.\arabic{equation}}

In this Appendix we briefly analyse the matching equations across the bubble wall 
(\ref{eq:vvs0}) for a symmetric bubble in the general case where the pressure is given by $p = - \mathcal{F}(\phi,T)$,
and describe the relation between the latent heat and the change in the relativistic degrees of freedom.
We start by decomposing $\mathcal{F}$ into

\be
\mathcal{F}(\phi,T) = \mathcal{F}(0,T) + \overline{\mathcal{F}}(\phi,T)\,,
\ee

\noindent where $\overline{\mathcal{F}}$ automatically fulfills $\overline{\mathcal{F}}(0,T) \equiv 0$. Furthermore, we can define
$\mathcal{F}(0,T = 0) \equiv \epsilon$ and $f(T) \equiv \mathcal{F}(0,T) - \mathcal{F}(0,T = 0) = -(a/3)\, T^4$, so that

\be
\mathcal{F}(\phi,T) = \epsilon - \frac{a}{3}\, T^4 + \overline{\mathcal{F}}(\phi,T)\,,
\ee

\noindent where $a$ is related to the number of relativistic degrees of freedom of the plasma
in the symmetric phase (it corresponds to the previously defined $a_+$). Then, the pressures and energy densities
in the symmetric and broken phases ($p_s, p_b, e_s$ and $e_b$) can be written as

\bea
\label{matchingsymmetric1}
p_b = \frac{a}{3}\, T_b^4 - \epsilon - \overline{\mathcal{F}}\left(\phi(T_b),T_b\right) \quad \quad \quad \quad 
p_s = \frac{a}{3}\, T_s^4 - \epsilon \nonumber\\
e_b = a\, T_b^4 + \epsilon + \overline{\mathcal{F}}\left(\phi(T_b),T_b\right) - T_b \, 
\left.\frac{\partial \overline{\mathcal{F}}(\phi,T)}{\partial T}\right|_{T_b} \quad \quad \quad \quad e_s = a\, T_s^4 + \epsilon 
\eea
We now define

\bea
\label{matchingsymmetric2}
\alpha(T) = \frac{1}{a \, T^4} \overline{\mathcal{F}}\left(\phi,T\right)\,, \quad \quad \quad \, \,\nonumber\\
\lambda(T) = \frac{1}{a \, T^4} \left[ \overline{\mathcal{F}}\left(\phi,T\right) - 
T \, \frac{\partial \overline{\mathcal{F}}(\phi,T)}{\partial T} \right]\,, \nonumber\\
r = \frac{T_b^4}{T_s^4} \,.\quad \quad \quad \quad \quad \quad \quad
\eea
$\alpha(T)$ is the ratio between the vacuum energy difference (between the broken and symmetric minima) and the thermal energy
of the plasma in the symmetric phase, and $\lambda(T)$ is the liberated latent heat. For the case of symmetric bubbles the matching
equations across the bubble wall (\ref{eq:vvs0}) can be written as

\be
\label{matchingsymmetric3}
v_b v_s = \frac{1- r \, \left(1-3\,\alpha(T_b)\right)}{3-3\, r \left(1+\lambda(T_b)\right)} \quad \quad
\frac{v_b}{ v_s} = \frac{3 + r \,\left(1-3\,\alpha(T_b)\right)}{1+3\,r\,\left(1+\lambda(T_b)\right)} 
\ee
For the nucleation of symmetric bubbles,
$T > T_c$ and so $\alpha(T)$ as defined in (\ref{matchingsymmetric2}) is positive

\be
\overline{\mathcal{F}}\left(\phi(T),T\right) -  \overline{\mathcal{F}}\left(0,T\right) > 0  \quad \longrightarrow \quad \alpha(T) > 0
\ee
However, since the variation of $\overline{\mathcal{F}}$ is the same as for the
usual symmetric-to-broken phase tunneling (the fact that now tunneling happens in the opposite direction does not change
the variation of $\overline{\mathcal{F}}$ with temperature), we actually have

\be
\frac{\partial \left[ \overline{\mathcal{F}}(\phi(T),T) - \overline{\mathcal{F}}(0,T) \right] }{\partial T} >
\frac{\overline{\mathcal{F}}(\phi(T),T) - \overline{\mathcal{F}}(0,T) }{T} > 0
\ee

\noindent which means that, as opposed to the usual case of broken bubbles,
for symmetric bubbles the latent heat $\lambda(T)$ is negative.
This is however related to the fact that, when converting broken to symmetric phase, the number of relativistic degrees
of freedom increases (increasing the thermal energy of the system). Comparing (\ref{matchingsymmetric1}) with the equivalent expressions
for the Bag EoS ((\ref{eosbag1}) and (\ref{eosbag2})), we get

\be
e_b = a\, T_b^4 + \epsilon + a\, T_b^4 \, \lambda(T_b) = a \left(1 + \frac{\epsilon}{a\, T_b^4} + \lambda(T_b)\right)  T_b^4
\equiv a_{-}\, T_b^4 
\ee

\noindent and it follows that in order to have $a_{-} < a \equiv a_{+}$, it is needed that $\lambda(T_b) < 0$. At the same time, 
the increase in the number of relativistic degrees of freedom favors the bubble expansion, since it contributes to the net 
pressure on the wall.

\section*{B. $\ $Symmetric Bubble Growth and $\delta r_0$}
\addcontentsline{toc}{section}{B $\ $Symmetric Bubble Growth and $\delta r_0$.}
\setcounter{equation}{0}
\renewcommand{\theequation}{B.\arabic{equation}}

\begin{figure}[ht]
\begin{center}
\includegraphics[width=0.65\textwidth, clip ]{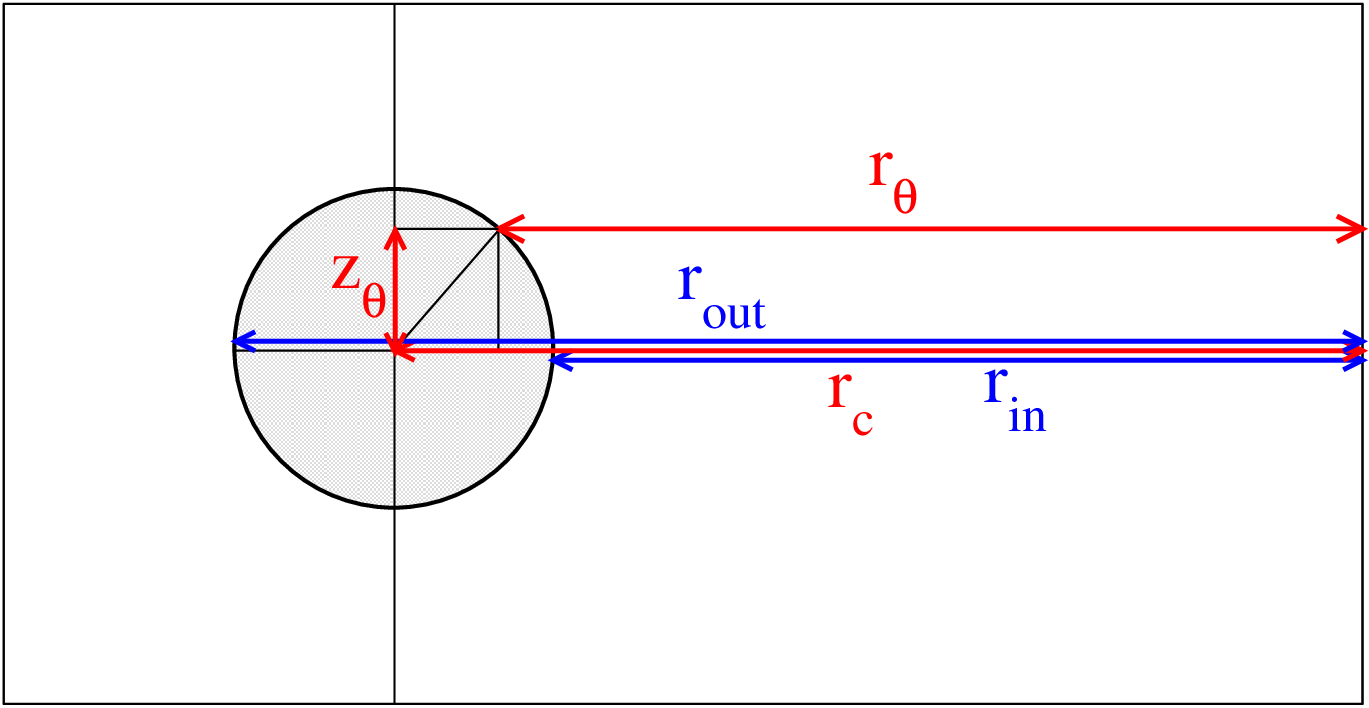}
\caption{\label{fig:14}
\small Parametrization of the symmetric bubble evolution.}
\end{center}
\end{figure}

As explained in section \ref{sec_5}, due to the detonation wave's inhomogeneous background
the symmetric bubble will be deformed from its initial spherical configuration as it expands. 
Neglecting the curvature of the broken bubble, the evolution of the symmetric bubble can be 
described by the variation of $r_{\theta}$ and $z_{\theta}$ (see Figure \ref{fig:14}) 

\be
\label{B1}
\frac{d\,r_{\theta}}{d\,t} = 
\frac{v\left(\frac{r_{\theta}}{t}\right) + \delta v \left(\frac{r_{\theta}}{t}\right) \mathrm{Sin}(\theta)}
{1 + v\left(\frac{r_{\theta}}{t}\right)\, \delta v \left(\frac{r_{\theta}}{t}\right) \mathrm{Sin}(\theta)} \quad \quad
\frac{d\,z_{\theta}}{d\,t} = \delta v \left(\frac{r_{\theta}}{t}\right) \mathrm{Cos}(\theta)
\, .
\ee

\be
\label{B2}
\mathrm{Sin}(\theta) = \frac{r_{\theta} - r_{c}}{r_{\frac{\pi}{2}} - r_{c}} \,\,\,\,\, \mathrm{for} \,\,\, \,\, \theta > 0 \quad\quad\quad 
\mathrm{Sin}(\theta) = - \frac{r_{c} - r_{\theta}}{r_{c} - r_{\frac{-\pi}{2}}} \,\,\,\,\, \mathrm{for} \,\,\,\,\,\theta < 0 
\, .
\ee

where $\delta v(\xi)$ is given by (\ref{eq:growth1}). Also, from Figure \ref{fig:14}
one has $r_{\frac{-\pi}{2}} \equiv r_{\mathrm{in}}$ and $r_{\frac{\pi}{2}} \equiv r_{\mathrm{out}}$ as 
defined in (\ref{eq:growth3}), and it is then clear that (\ref{eq:growth3}) derives from (\ref{B1}). 

\begin{figure}[ht]
\begin{center}
\includegraphics[width=0.8\textwidth, clip ]{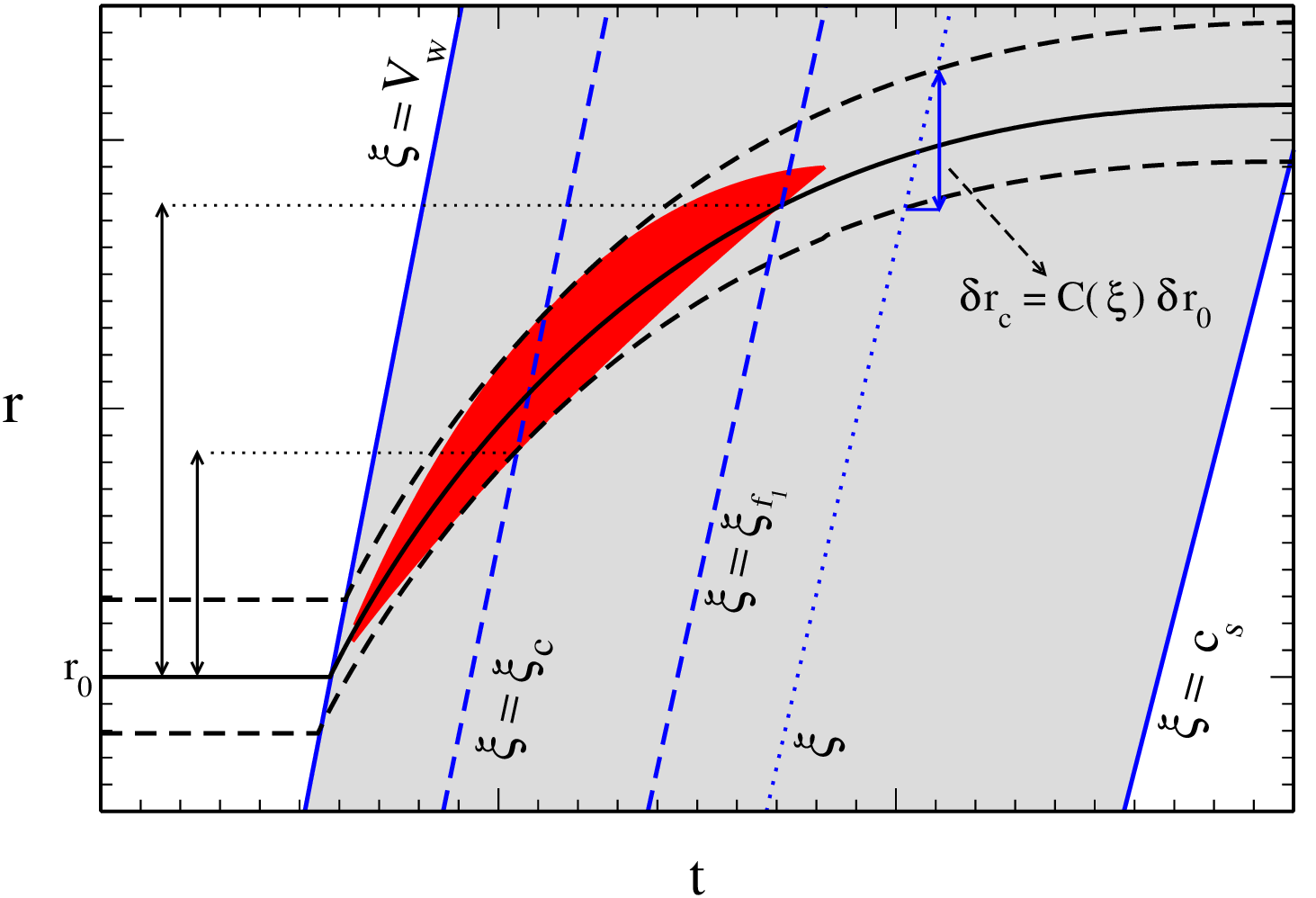}
\caption{\label{fig:15}
\small
Detail of Figure \ref{fig:5}, where the scale-invariant (self-similar) ratios used to 
prove (\ref{eq:detonationelement2}) and the $r_0$-independence of (\ref{B5}) and (\ref{B6}) are shown explicitly.}
\end{center}
\end{figure}

Let us now turn to the evaluation of the volume of plasma that goes 
through a symmetric bubble. Focusing on the broken bubble's radial direction, 
this amounts to obtain the quantities $\delta r_0^{(2)}$ and $\delta r_0^{(1)}$ as a function of $r_0$ and $\xi$ (as defined
in (\ref{eq:growth5})). This can be done using the self-similarity of the 
detonation wave. We start by defining $\xi_{f_1}$ as (see Figure \ref{fig:15})  

\be
\label{B3}
r_{\mathrm{in}}(r_0, \xi_{f_1}) = r_c(r_0, \xi_{f_1})  
\, .
\ee
Then, due to the self-similarity of the detonation wave, and using (\ref{eq:detonationelement2}) and (\ref{B3}), the ratio

\be
\label{B4}
\frac{r_{\mathrm{in}}(r_0, \tilde{\xi})-r_0}{r_{\mathrm{in}}(r_0, \xi_{f_1})-r_0} = 
\frac{r_{\mathrm{in}}(r_0, \tilde{\xi})-r_0}{r_c(r_0, \xi_{f_1}) -r_0} \simeq 
\frac{1}{C(\xi_{f_1}) -1}  \frac{r_{\mathrm{in}}(r_0, \tilde{\xi})-r_0}{r_0} 
\, ,
\ee
does not depend on the value of $r_0$ (see Figure \ref{fig:15}). Since $C(\xi_{f_1}) -1$ is already independent of $r_0$, we get

\be
\label{B5}
\frac{r_{\mathrm{in}}(r_0, \tilde{\xi})}{r_0} \equiv K_{1}(\tilde{\xi})   \quad \quad \quad \quad (r_0\mathrm{-independent})
\, .
\ee
Using similar arguments, one also has 

\be
\label{B6}
\frac{r_{\mathrm{out}}(r_0, \tilde{\xi})}{r_0} \equiv K_{2}(\tilde{\xi})   \quad \quad \quad \quad (r_0\mathrm{-independent})
\, .
\ee
Then, from (\ref{eq:detonationelement2}) and (\ref{eq:growth5}) we obtain 
 
\be
\label{B7}
\delta r_0^{(1)} = r_0 \left( 1- \frac{K_{1}(\tilde{\xi}) }{C(\tilde{\xi})} \right)
\quad \quad \quad 
\delta r_0^{(2)} = r_0 \left(\frac{K_{2}(\tilde{\xi})}{C(\tilde{\xi})}  -1 \right)
\, .
\ee
Finally, we arrive at

\be
\label{B8}
\delta r_0 \equiv \delta r_0^{(1)} + \delta r_0^{(2)} = r_0 
\left( \frac{K_{2}(\tilde{\xi}) - K_{1}(\tilde{\xi}) }{C(\tilde{\xi})}\right)
\, .
\ee

As a final remark, even though
(\ref{B1}) can be used to study the deviation from spherical symmetry of the symmetric bubbles as they expand,
for all practical purposes concerning the mechanism of EW baryogenesis outlined in section \ref{sec_6}
one can neglect its effects on the computation of the volume of plasma that goes through a symmetric bubble $V_S$
and assume (\ref{eq:growth6})

\be
\label{B9}
V_S \simeq (\delta r_0)^3
\, .
\ee

\end{document}